\documentclass[prd,aps,twocolumn,a4paper,floatfix,showpacs,nofootinbib,superscriptaddress]{revtex4-1}

\usepackage{times}
\usepackage[usenames]{color}
\usepackage{amsfonts}
\usepackage{siunitx}
\usepackage{amsmath}
\usepackage{amssymb}
\usepackage{bm}
\usepackage{dcolumn}
\usepackage{enumerate}
\usepackage{epsfig}
\usepackage{graphicx}
\usepackage{tabularx}
\usepackage{graphics}
\usepackage{gensymb}
\usepackage{multirow}
\usepackage[latin1]{inputenc}
\usepackage{latexsym}
\usepackage{rotating}
\usepackage{hyperref}
\usepackage{nicefrac}
\usepackage[normalem]{ulem}
\usepackage[caption=false]{subfig}
\usepackage{autobreak}
\usepackage{textcomp}
\usepackage{siunitx}
\usepackage{float}
\definecolor{orange}{rgb}{0.9,0.5,0}
\hypersetup{
    colorlinks=true,
    linkcolor=red,
    filecolor=magenta,      
    urlcolor=magenta,
    citecolor=magenta,
}

\graphicspath{ {./figures/} }

\definecolor{mgreen}{rgb}{0.1,0.7,0.1}

\begin{document}
\title{Conformally curved initial data for charged, spinning black hole binaries on arbitrary orbits}

\author{Soham~Mukherjee}
\affiliation{Department of Physics and Astronomy, University of Waterloo, Waterloo, Ontario, Canada, N2L 3G1}
\affiliation{Perimeter Institute for Theoretical Physics, 31 Caroline Street North, Waterloo, Ontario, Canada, N2L 2Y5}
\author{Nathan~K.~Johnson-McDaniel}
\affiliation{International Centre for Theoretical Sciences, Tata Institute of Fundamental Research, Bengaluru 560089, India}
\affiliation{Department of Applied Mathematics and Theoretical Physics, Centre
for Mathematical Sciences, University of Cambridge, Wilberforce Road,
Cambridge, CB3 0WA, UK}
\affiliation{Department of Physics and Astronomy, University of Mississippi, University, Mississippi 38677, USA}
\author{Wolfgang~Tichy}
\affiliation{Department of Physics, Florida Atlantic University, Boca Raton, Florida 33431 USA}
\author{Steven~L.~Liebling}
\affiliation{Long Island University, Brookville, New York 11548, USA}
\date{\today}

\begin{abstract}
We present a method to construct conformally curved initial data for charged black hole binaries with spin on arbitrary
orbits. We generalize the superposed Kerr-Schild, extended conformal thin sandwich construction from
[Lovelace~\emph{et~al.}, Phys.~Rev.~D {\bf{78}}, 084017 (2008)] to use Kerr-Newman metrics for the superposed black holes
and to solve the electromagnetic constraint equations. We implement the construction in the pseudospectral code SGRID. The
code thus provides a complementary and completely independent excision-based construction, compared to the existing
charged black hole initial data constructed using the puncture method [Bozzola and Paschalidis, Phys.~Rev.~D {\bf{99}},
104044 (2019)]. It also provides an independent implementation (with some small changes) of the Lovelace~\emph{et~al.}\
vacuum construction. We construct initial data for different configurations of orbiting binaries, e.g., with black holes
that are highly charged or rapidly spinning (90 and 80 percent of the extremal values, respectively, for this initial
test, though the code should be able to produce data with even higher values of these parameters using higher
resolutions), as well as for generic spinning, charged black holes. We carry out exploratory evolutions with the finite
difference, moving punctures codes BAM (in the vacuum case) and HAD (for head-on collisions including charge), filling
inside the excision surfaces. In the charged case, evolutions of these initial data provide a proxy for binary black hole
waveforms in modified theories of gravity. Moreover, the generalization of the construction to Einstein-Maxwell-dilaton
theory should be straightforward.
\end{abstract}
\maketitle

\section{Introduction}
\label{sec:intro} 
Gravitational wave~(GW) observations of compact binary mergers, primarily binary black holes (e.g.,~\cite{LIGOScientific:2018mvr, Abbott:2020niy,LIGOScientific:2021djp}), have made it possible to test general
relativity~(GR) in the strong-field, high velocity regime where GR is most likely to break down (see~\cite{Abbott:2018lct,LIGOScientific:2020tif,LIGOScientific:2021sio} for results of such tests carried out by
the LIGO-Virgo collaboration). However, these tests are all null tests of one sort or another, and one would ideally want to compare the predictions of GR for the coalescence of compact
binaries with the predictions of a suite of well-motivated alternative theories (in particular, \cite{Chua:2020oxn} discusses some of the problems encountered with certain null tests). To carry out such comparisons, one needs to construct high-accuracy waveform models in alternative theories, just as have been constructed in GR. At present, constraints on alternative theories using binary black hole observations have been restricted to using model waveforms constructed with only low-order post-Newtonian (PN) calculations in the alternative theories~\cite{Perkins:2021mhb,Wang:2021yll}.

Numerical relativity simulations of the late inspiral and merger in alternative theories will be a key ingredient in constructing such waveform models, just as in GR. However, such
simulations come with several challenges. Many of these modified theories either do not have a \emph{known} well-posed initial value formulation (making them unsuitable for numerical
evolution) or lack a construction of constraint-satisfying initial data for compact binaries. While significant progress has been made towards finding a well-posed formulation for
Einstein-scalar-Gauss-Bonnet, Lovelock, and Horndeski theories at weak coupling~\cite{Kovacs:2020pns, Kovacs:2020ywu} and their subsequent numerical
evolution~\cite{East:2020hgw,East:2021bqk,Figueras:2021abd}, most approaches to simulating binary black holes in modified theories of gravity depend on an order reduction method~\cite{Okounkova:2019zjf, Okounkova:2019dfo, Okounkova:2020rqw, Witek:2018dmd,Silva:2020omi}. Such approaches compute the effects of modified theories as perturbations to the GR solution, which leads to a secular drift between the true solution and the perturbative solution~\cite{Okounkova:2019zjf, Okounkova:2020rqw}, though there are proposals for methods to remove this drift~\cite{GalvezGhersi:2021sxs}. There are also approaches that modify the equations to make the theories well-posed~\cite{Cayuso:2020lca}, but not yet any evolutions of binary black holes with such approaches.

For theories that do have a known well-posed initial value formulation, evolutions of binary black hole mergers have been
carried out using approximate initial data. Examples include a study~\cite{Hirschmann:2017psw} in Einstein-Maxwell-dilaton
theory~\cite{Gibbons:1987ps} and evolutions in scalar-tensor theories of gravity~\cite{Berti:2013gfa,Healy:2011ef} where,
in the absence of externally imposed scalar field dynamics, binary black holes in scalar-tensor theory have the same
dynamics as those in vacuum GR.

Here we consider charged binary black holes in Einstein-Maxwell theory. This provides a well-posed framework that mimics some of the
features of binary black hole mergers in modified theories. Additionally, various modified theories besides Einstein-Maxwell-dilaton theory
also contain vector fields with a Maxwell-like kinetic term (see, e.g.,~\cite{Will:2014kxa}), so this is a first step to performing simulations in those theories.
The specific features of Einstein-Maxwell theory that mimic some of the effects seen in modified theories are modified or additional PN terms in the
dynamics of the binary (see, e.g.,~\cite{Khalil:2018aaj}) and differences
in the spectrum of quasinormal modes for the final black hole (see, e.g.,~\cite{Dias:2015wqa,Dias:2021yju,Carullo:2021oxn}). Both of these effects are directly
encoded in the binary's GW signal. In particular,
charged binary black holes with unequal charge-to-mass ratios emit dipole radiation (see, e.g.,~\cite{Khalil:2018aaj}), a
common feature in several modified theories of gravity, e.g., scalar Gauss-Bonnet~(sGB) gravity~\cite{Shiralilou:2021mfl}
and scalar-tensor theories of gravity~\cite{Sennett:2016klh}. Binary black holes emit dipole radiation due to a charge in
sGB gravity, as well, though there it is a scalar charge, as opposed to the $\rm{U}(1)$ charge we consider here. Binary
black holes in scalar-tensor theories do not emit dipole radiation, but systems with matter will in general emit scalar
dipole radiation in those theories. Nevertheless, the leading PN effect of the dipole radiation on the binary's dynamics
will have the same frequency dependence in all cases, viz., a $-1$PN order contribution to the orbital phasing and thus to
the phase of the GW signal (in the case where the dipole radiation can be treated as a perturbation to the
dominant quadrupole radiation---see, e.g.,~\cite{Khalil:2018aaj}). 

From an observational perspective, the LIGO-Virgo analyses (e.g.,~\cite{Abbott:2018lct,LIGOScientific:2020tif,LIGOScientific:2021sio}) currently test for the presence of such an additional $-1$PN term in the
phase. In the absence of dipole radiation (which is the case for charged binary black holes with equal charge-to-mass ratio and the same sign of charge) the deviations from vacuum GR will
only occur at $1\text{PN}$ order and above (since the $0\text{PN}$ modifications are degenerate with a rescaling of the binary's masses, as discussed in, e.g.,~\cite{Cardoso:2016olt}).
This is similar to some modified theories, e.g., dynamical Chern-Simons theory, where the deviations from GR start at $2\text{PN}$~\cite{Yagi:2011xp}. Many of the other LIGO-Virgo tests of general
relativity check (explicitly or implicitly) for such deviations in higher-order PN coefficients. Additionally, a recent analysis~\cite{Carullo:2021oxn} checks for the presence of charge in
the ringdown waveforms, but finds that one can only place very weak constraints using the ringdown phase alone due to correlations between the charge and spin parameters.

Charged binary black hole waveforms, therefore, provide a proxy for BBH waveforms from modified theories of GR that would
allow us to test the sensitivity of current LIGO-Virgo tests of GR to completely consistent, parameterized deviations from
GR (see~\cite{Johnson-McDaniel:2021yge} for an initial study using phenomenological waveforms). Since the charge can be
varied freely up to the extremal limit, this allows for significant deviations from vacuum GR. Further, one can use the
waveforms from numerical simulations of charged binary black holes to create a waveform model that one can use to analyze
gravitational wave data. Such a waveform model would likely also have significant input from PN
calculations~\cite{Khalil:2018aaj,Julie:2018lfp} and black hole perturbation theory computations of quasinormal
modes~\cite{Dias:2015wqa,Dias:2021yju,Carullo:2021oxn}, and possibly also from black hole perturbation theory/self-force
calculations of waveforms in the extreme mass-ratio limit (see \cite{Zhu:2018tzi,Burton:2020wnj,Torres:2020fye} for work
in this direction involving charge). Such a model would allow one to constrain the charges of observed black holes. Such
an analysis has already been carried out for low-mass, inspiral-dominated signals in~\cite{Gupta:2021rod} using a
simple waveform model created by combining a vacuum GR model with the known, still relatively low-order PN contributions
from the charges from~\cite{Khalil:2018aaj}. There is also a study of GW150914 using numerical relativity waveforms of
charged binary black holes and simple data analysis arguments in~\cite{Bozzola:2020mjx} (see also further data analysis
calculations with these waveforms in~\cite{Bozzola:2021elc}).

A full waveform model tuned to numerical relativity simulations would also allow one to constrain the charges of high-mass
binaries, as well as firming up the constraints obtained with the simple waveform model presented in~\cite{Gupta:2021rod}.
While astrophysical black holes are expected to have negligible electric charges (as reviewed in~\cite{Cardoso:2016olt}),
they could have nonnegligible magnetic charges if magnetic monopoles exist (as discussed
in~\cite{Liebling:2016orx,Preskill:1984gd}). Additionally, black holes could attain a significant charge in the scenario
where the black holes are charged under a dark electromagnetic field, e.g., the minicharged dark matter scenario
considered in~\cite{Cardoso:2016olt}. As a first step towards constructing such waveform models, we construct initial data
for spinning, charged binary black holes in orbit. Numerical relativity simulations of charged binary black hole mergers
also provide the mapping from the initial masses, spins, and charges to the final mass and spin, and better knowledge of
this mapping, particularly for close-to-extreme charges, will improve the modeling of possible gamma-ray backgrounds from
mergers of charged primordial black holes~\cite{Kritos:2021nsf}.

Several previous studies have investigated the dynamics of charged black holes in full numerical relativity.
Zilh{\~a}o~\emph{et al.}~\cite{Zilhao:2012gp, Zilhao:2013nda} performed simulations of head-on collisions of nonspinning
charged black holes from rest, using either analytic equal charge-to-mass ratio initial data or a simple numerical initial
data construction to obtain opposite charge-to-mass ratios. Liebling~\emph{et al.}~\cite{Liebling:2016orx} carried out
evolutions for weakly charged (electric and magnetic) black holes on quasicircular orbits starting from approximate
initial data. Most notably, Bozzola and Paschalidis~\cite{Bozzola:2019aaw} constructed conformally flat, puncture initial
data for multiple charged black holes with linear and angular momenta using the conformal transverse traceless approach,
and subsequently evolved a set of nonspinning binary black holes with small to moderate charges on quasicircular orbits
in~\cite{Bozzola:2020mjx, Bozzola:2021elc} and, more recently, head-on collisions of boosted, charged black holes in~\cite{Bozzola:2022uqu}.

We take a different approach to solving for charged binary black hole initial data. Specifically, we construct conformally
curved excision initial data by extending the approach developed by Lovelace \emph{et~al.}\ in~\cite{Lovelace:2008tw} to
include charge.\footnote{There are improvements to this approach in~\cite{Varma:2018sqd, Ma:2021can, Chen:2021rtb}, but we
follow the original construction.} The approach is based on the Extended Conformal Thin-Sandwich
formalism~\cite{York:1998hy,Pfeiffer:2002iy} and uses the superposition of two boosted Kerr black holes in Kerr-Schild
coordinates (weighted by attenuation functions) for the conformal metric. Compared to the conformal transverse traceless
construction, this approach has several advantages (at least in the uncharged case). For example it allows for higher
spins~\cite{Lovelace:2008tw}, less junk radiation~\cite{Lovelace:2008hd}, and better control over the physics through the
boundary conditions at the excision surfaces. For our case, we replace the Kerr black holes by Kerr-Newman black holes and
solve for the final electric field by solving for a correction to the superposed electric field (given by a scalar
potential) to satisfy the divergence constraint. We use a simple boundary condition for the scalar potential on the
excision surfaces that allows us to fix the charges of the black holes. We compute the magnetic field from the superposed
vector potentials so that it satisfies the divergence constraint by construction.

In this paper, we generate excision based initial data for both highly charged and highly spinning binaries using the
pseudospectral initial data solver SGRID~\cite{Tichy:2009yr,Dietrich:2015pxa}, and perform exploratory evolutions of the
initial data in some cases with both the BAM~\cite{Brugmann:2008zz} and HAD~\cite{Liebling:2020jlq} evolution codes. Both
of these codes use finite differences and moving punctures for evolution, but they use different algorithms to fill inside
the excision surfaces, and only HAD can evolve the charged case.

In Sec.~\ref{sec:id}, we first review the construction from~\cite{Lovelace:2008tw} and then give its extension to the
charged case. In Sec.~\ref{sec:num}, we discuss the details of our numerical implementation. We then show examples of our
initial data construction and exploratory evolutions using BAM and HAD in Sec.~\ref{sec:results}. We conclude in
Sec.~\ref{sec:discussions}. In Appendix~\ref{append:KS}, we give expressions for a Kerr-Newman black hole in Kerr-Schild
form, while in Appendices~\ref{append:BHFiller} and~\ref{append:BHFillerHAD}, we give details of the black hole filling
algorithms used in BAM and HAD, respectively. We use lowercase Latin letters to denote spatial indices and Greek letters
for spacetime indices. We reserve the index $A$ to label the black holes. Finally, we use geometric units $(G = c = 1)$
and Gaussian units for the electromagnetic field throughout the paper.

\section{Initial Data Formalism}
\label{sec:id}
To compute constraint satisfying initial data on a Cauchy hypersurface, one needs to solve the Hamiltonian and momentum
constraints for the geometry and the electomagnetic constraint equations. With standard decompositions, these constraints
constitute a set of coupled, nonlinear, second order differential equations
for a given set of freely specifiable variables, and this set has to be supplemented with appropriate boundary conditions; see,
e.g.,~\cite{Baumgarte:2002jm} for an introduction to the gravitational constraint equations. There are several approaches to solving this problem, each with a
different decomposition of the gravitational constraint equations and a corresponding unique group of freely specifiable variables (see,
e.g.,~\cite{Cook:2000vr, Tichy:2016vmv} for reviews). Different decompositions can result in different initial data, e.g.,
with different amounts of junk radiation~\cite{Cook:2000vr, Alc08} that could lead to different physical parameters of
the binary. We now review one such decomposition, the Extended Conformal Thin Sandwich formalism, which forms the basis of
our initial data construction.

\subsection{The Extended Conformal Thin Sandwich Formalism}
\label{subsec:XCTS} 
The Extended Conformal Thin Sandwich (XCTS) formalism~\cite{York:1998hy, Pfeiffer:2002iy} is an alternative approach to
the transverse-traceless construction~\cite{Bowen:1980yu} for calculating initial data. In contrast to the
transverse-traceless construction, it allows for some degree of control over the time evolution of the initial data. As is standard, this construction begins by splitting the spatial metric $\gamma_{ij}$ into a conformal factor $\psi$
and the conformally related metric $\tilde\gamma_{ij}$, and then splitting the extrinsic curvature $K_{ij}$ into its trace $K$
and a traceless part $A_{ij}$, giving
\begin{subequations}
\begin{align}
	\gamma_{ij} &= {} \psi^4 \tilde{\gamma}_{ij}, \\
	K_{ij} &= {} A_{ij} + \dfrac{1}{3} \gamma_{ij} K. 
\end{align}
\end{subequations}
In the XCTS formalism, we are allowed to freely specify $\{\tilde{\gamma}_{ij}, K\}$ together with their time derivatives
$\{\tilde{u}_{ij}, \partial_t K\}$. We then solve for $\{\gamma_{ij}, K_{ij}\}$ in terms of $\psi$, the shift vector $\beta_i$,
and the lapse $\alpha$ multiplied by the conformal factor ($\alpha\psi$), respectively. 

\subsubsection{XCTS equations}
For a given choice of freely
specifiable variables, the Hamiltonian and momentum constraint equations decompose into a set of coupled
elliptic equations for $\psi, \alpha\psi$, and $\beta_i$ given by
\begin{subequations}
\label{eq:xcts}
\begin{widetext}
\begin{align}
	\tilde\nabla_{j}\tilde\nabla^{j}\psi - \dfrac{1}{8}\tilde R\psi - \dfrac{1}{12}K^{2}\psi^{5} + 
	\dfrac{1}{8}\psi^{-7}\tilde A^{ij} \tilde A_{ij}  &= -2\pi\psi^{5}\rho, \\
	\tilde\nabla_{j}\left[\dfrac{\psi^{7}}{2(\alpha\psi)} \left(\tilde{L}\beta\right)^{ij}\right]
	- \dfrac{2}{3}\psi^{6}\tilde\nabla^{i}K
	- \tilde\nabla_{j}\left[ \dfrac{\psi^{7}}{2(\alpha\psi)} \tilde u^{ij}\right] &= 8\pi\psi^{10}J^{i}, \\
	\tilde\nabla_{j}\tilde\nabla^{j}(\alpha\psi) - (\alpha\psi)\left[\dfrac{\tilde R}{8} + \dfrac{5}{12}K^{2}\psi^{4} 
	+ \dfrac{7}{8}\psi^{-8}\tilde A^{ij} \tilde A_{ij} \right] +
	\psi^{5}(\partial_{t}K - \beta^{k}\partial_{k}K) &= (\alpha\psi)\left[2\pi\psi^{4}(\rho + 2\sigma)\right],
\end{align}
\end{widetext}
\end{subequations}
where
\begin{subequations}
\begin{align}
	(\tilde L \beta)_{ij} := \tilde\nabla_{i}\beta_{j} +  \tilde\nabla_{j}\beta_{i} - 
	\dfrac{2}{3}\tilde \gamma_{ij}\tilde\nabla_{k}\beta^{k}, \\
	 \tilde A^{ij} = \psi^{10}A^{ij} = \dfrac{\psi^{7}}{2(\alpha\psi)} 
	 \left[(\tilde L\beta)^{ij} - \tilde u^{ij}\right].
\end{align}
\end{subequations}
In these equations, the terms decorated with a tilde are associated with the conformal metric. Thus,
$\tilde{\nabla}_{j}$ and $\tilde{R}$ are the covariant derivative and Ricci scalar associated with
$\tilde{\gamma}_{ij}$, respectively. We also have electric ($E_i$) and
magnetic ($B_i$) fields due to the presence of charged black holes, so the expressions for energy density $\rho$, the momentum
density $J^i$, and the trace of the stress tensor as measured by an Eulerian observer $\sigma$ are given
by~\cite{Alcubierre:2009ij}
\begin{subequations}
\begin{align}
	\rho = {} & \dfrac{1}{8\pi}\left(E^{i}E_{i} + B^{i}B_{i}\right) = \sigma, \\
	\label{eq:J}
	J^{i} = {} & \dfrac{1}{4\pi} (\mathbf{E} \times \mathbf{B})^i, 
\end{align}
\end{subequations}
where
\begin{equation}
         \label{eq:cross_product_curved_space}
	(\mathbf{E} \times \mathbf{B})^i = \dfrac{1}{\sqrt{\gamma}}\,\epsilon_F^{ikl} E_k B_l
\end{equation}
is the curved-space version of the cross product. Here $\gamma$ is the determinant of $\gamma_{ij}$ and $\epsilon_F^{ikl} = \epsilon^F_{ikl}$ is the flat space Levi-Civita symbol taking values in $\{0, \pm 1\}$. Here, and in the rest of the paper, we raise and lower indices of the physical fields, e.g., $E_i$ and $B_i$, using the physical metric. For the conformally related variables, we use the conformal metric to raise and lower indices. We use boldface letters to denote three-dimensional vector quantities.

\subsubsection{Choice of freely specifiable variables}
Given the XCTS equations~\eqref{eq:xcts}, the next step is to make a choice for the freely specifiable variables $\{\tilde{\gamma}_{ij}, K\}$ and $\{\tilde{u}_{ij}, \partial_t K\}$. For
$\tilde{\gamma}_{ij}$ and $K$, we closely follow the construction by Lovelace~\emph{et al.}~\cite{Lovelace:2008tw}, and superpose two boosted, spinning Kerr-Newman black holes in
Kerr-Schild coordinates (see Appendix~\ref{append:KS} for the complete expressions) weighted by Gaussian attenuation functions centered around each black hole. Specifically, we set
\begin{subequations}
		\label{eq:sks}
\begin{align}
	\tilde{\gamma}_{ij} = {} & f_{ij} + \sum_{A=1}^2 e^{-r_{A}^2 / w_{A}^2} (\gamma^{A}_{ij} - f_{ij}), \\
	K = {} & \sum_{A=1}^2 e^{-r_{A}^2 / w_{A}^2} K_{A}.
\end{align}
\end{subequations}
Here $\gamma^{A}_{ij}$ and $K_{A}$ are the spatial metric and the trace of the extrinsic curvature of black hole $A$ where $A \in \{1,2\}$, and $f_{ij}$ is the flat space spatial metric.
Further, $r_{A}$ and $w_{A}$ are the coordinate distance from the center of each black hole and the freely specifiable attenuation weight for that black hole, respectively. The free
parameters $w_{A}$ allow one to control the extent of influence of each black hole on the other. In particular, these attenuation functions limit significant deviations from maximal slicing assumption ($K=0$) and conformal flatness to the regions surrounding each black hole.\footnote{While having asymptotically conformally flat data was shown to be necessary for consistency of the outer boundary conditions (and thus to obtain exponential convergence of the initial data) in~\cite{Lovelace:2008hd}, the same argument does not apply to our construction, since we do not put the corotation term in the outer boundary condition (see Sec.~\ref{sec:xctsbc}) as was done in~\cite{Lovelace:2008hd}.} The choice of $w_{A}$ affects the amount of
junk radiation, and can thus be adjusted to minimize the junk radiation, as mentioned in~\cite{Lovelace:2008hd}. We discuss our choices for these weights in Sec.~\ref{sec:results}.

We set the freely specifiable time derivatives, i.e., $\{\tilde{u}_{ij}, \partial_t K\}$, to zero, again as in~\cite{Lovelace:2008tw}. For binary black holes in quasicircular orbit, or for
eccentric orbits at apoapsis, this is a reasonable approximation, since we can expect the system to be in quasi-equilibrium in an instantaneously corotating frame. For highly boosted
head-on collisions, however, this approximation will break down with increasing boost, and we can no longer set these quantities to zero while still obtaining accurate initial data. We
will return to this point in Sec.~\ref{sec:results}.

\subsubsection{Boundary Conditions}
\label{sec:xctsbc}
The final components of our initial data construction for the geometry are the boundary conditions we need to impose on the domain boundaries.
Our numerical grid has two boundaries (discussed further in Sec.~\ref{sec:grid}). The outer boundary is located at spatial infinity ($i_0$)
and the inner boundaries are the excision surfaces $(\mathcal{S}_A)$ for the two black holes. At spatial infinity,
we again follow~\cite{Lovelace:2008tw} and set
\begin{subequations}
	\label{eq:bc_outer}
	\begin{align}
	\psi = {} & 1\quad\text{at}~i_0, \\
	\alpha\psi = {} & 1\quad\text{at}~i_0.
\end{align}
\end{subequations}
These equations ensure that our spatial metric is asymptotically flat. For $\beta^i$, we follow~\cite{Ossokine:2015yla} but remove the corotation and expansion terms ($\mathbf{\Omega_0} \times \mathbf{r}$ and $\dot{a}_0 r^i$) from the corresponding expression, and set
\begin{equation}
	\label{eq:bc_B_inf}
	\beta^k = v_0^k\quad\text{at}~i_0.
\end{equation}
Here $v_0^i$ is a velocity parameter used to drive the Arnowitt-Deser-Misner (ADM) linear momentum to zero in our initial data (see Sec.~\ref{subsec:iter}). The corotation and expansion terms diverge at spatial infinity, and thus cause problems in our numerical setup with a compactified grid. We introduced auxiliary variables to handle the diverging terms analytically, but were still unable to get our initial data solver to converge when including these terms in the outer boundary conditions. Instead, we transfer the corotation and expansion terms usually included in the outer boundary condition to the inner boundary condition on $\beta^i$. This transferral leads to equivalent initial data in the conformally flat case, as shown for the corotation term in~\cite{Pfeiffer:2007yz} but is not equivalent in the conformally curved case.\footnote{The same argument also holds for the expansion term, since it is also a conformal Killing vector in this case.}  However, we find that our modified boundary conditions for the shift also leads to orbiting black holes.

Concretely, we set
\begin{align}
	\label{eq:bc_shift_inner}
 	\beta^{i} &= \alpha s^{i} - \Omega_r^{k}\xi^{i}_{~(k)} - [\mathbf{\Omega}_{0}\times (\mathbf{r} - 	\mathbf{r}_\text{CM})]^{i}\nonumber\\ 
	 &\quad - \dot{a}_0 (\mathbf{r} - 	\mathbf{r}_\text{CM})^i\quad\text{on}~\mathcal{S}_A.
\end{align}
Here, we use the Euclidean cross product. Further, $s^i$ is the outward-pointing normal to the excision surface,
$\Omega_r^k$ is a
free parameter (similar to $v_0^i$) used to control the magnitude of the spin of each black hole as in, e.g.,~\cite{Lovelace:2008tw}, and $\xi^{i}_{~(k)}$ are approximate rotational
Killing vectors on $\mathcal{S}_A$. Additionally, $\mathbf{r}$ and $\mathbf{r}_{\text{CM}}$ correspond to the coordinate position vector and the position of the Newtonian center of mass of
the binary, respectively. Finally, $\mathbf{\Omega}_{0}$ and $\dot{a}_0$ correspond to the orbital and radial velocity of the binary, which are adjusted iteratively after evolving the initial data to reduce the
eccentricity of the binary. We discuss our eccentricity reduction procedure in Sec.~\ref{sec:evolutions}. Here $\Omega_r^k$ and $\xi^{i}_{~(k)}$ are in general different for the two black holes, but we omit the $A$
label on them, for notational simplicity. 

We take $\xi^{i}_{~(k)}$ to be the rotational Killing vectors in flat space, i.e.,
\begin{equation}
	\label{eq:Killingvec}
	\xi^{i}_{~(k)} = \epsilon^F_{ikl}\left(\mathbf{r} - \mathbf{r}_{\text{BH}}\right)^{l},
\end{equation}
the same choice used to measure black hole spins in BAM~\cite{Lages_thesis}, where $\mathbf{r}_{\text{BH}}$ is the coordinate location of the center of each black hole (again omitting the
$A$ label). Additionally, these flat space rotational Killing vectors lead to spin measurements that agree well with PN predictions for nutation~\cite{Owen:2017yaj}. Thus, the more
involved approximate Killing vector constructions in~\cite{Cook:2007wr, Lovelace:2008tw} may not necessarily lead to better initial data. In particular, it is likely most useful for
waveform modeling purposes to have spin measurements that agree well with the PN definitions. However, it would likely be worthwhile to consider the boost-fixed version 
of the flat space rotational Killing vectors
introduced in~\cite{Owen:2017yaj}.

For the rest of the inner boundary conditions, we follow~\cite{Lovelace:2008tw} and set
\begin{subequations}
\label{eq:bc_psi_alphapsi_inner}
\begin{align}
	\tilde{s}^{k}\partial_{k}\psi &= -\dfrac{\psi^{3}}{8\alpha}\tilde{s}^{i}\tilde{s}^{j}(\tilde{L}\beta)_{ij} - 	\dfrac{\psi}{4}\tilde{h}^{ij} \tilde{\nabla}_{i}\tilde{s}_{j} + \dfrac{1}{6}K\psi^{3}\quad\text{on}~\mathcal{S}, \\
	\alpha\psi 					  &=  1 +  \sum_{A=1}^2 e^{-r_{A}^2/w_{A}^2}(\alpha^{A} - 1)\quad\text{on}~\mathcal{S},
\end{align}
\end{subequations}
where $\tilde{s}^i$ and $\tilde{h}_{ij}$ are the conformally-rescaled surface
normal and the conformal 2-metric on $\mathcal{S} := \mathcal{S}_1 \cup \mathcal{S}_2$, respectively. (Here we just refer to $\mathcal{S}$, since these boundary conditions do not have anything specific to a given excision surface.) Specifically,
\begin{align}
	\label{eq:hij}
	h_{ij} = \gamma_{ij} - s_{i}s_{j} = \psi^{4}(\tilde{\gamma}_{ij} - \tilde{s}_{i}
	\tilde{s}_{j}) = \psi^{4}\tilde{h}_{ij}.
\end{align}
This inner boundary condition on the conformal factor ensures that the excision surfaces coincide with the apparent horizons, while the condition on the lapse is a gauge choice that ensures that the time coordinate near each black hole is close to that of the corresponding Kerr-Schild spacetime. 

\subsection{Solving the electromagnetic constraint equations}
\label{sec:excts}
For binary black hole spacetimes with electric charge, we will have nonzero electric and (in general) magnetic fields. These electromagnetic (EM) fields, like the geometric quantities in the previous section, cannot be freely specified on the Cauchy surface and need to satisfy the EM constraint equations
\begin{subequations}
\begin{align}
	\label{eq:divErho}
	\nabla_i E^i = {} & 4\pi\mathbf{\rho_{\mathrm{EM}}}, \\ 
	\label{eq:divB}
	\nabla_i B^i = {} & 0.
\end{align}
\end{subequations}
(See~\cite{Alcubierre:2009ij} for the $3+1$ decomposition of the Einstein-Maxwell equations.)
Here $\rho_\mathrm{EM}$ is the charge density as measured by an Eulerian observer and $\nabla_i$ is the covariant derivative compatible with the physical metric $\gamma_{ij}$. Similar to the XCTS equations, we choose a particular decomposition of these equations, which uniquely determines the degrees of freedom we solve for and the choice of freely specifiable variables. 

To solve for the electric field, we start by introducing a correction to the background (superposed) electric field $(E_{\text{sp}})^i$ in the form of a gradient of a scalar potential\footnote{A different method to construct a constraint satisfying electric field would be to superpose the individual electric fields weighted by the determinants of the individual
physical $3$-metrics as suggested by East~\cite{East:2018glu}, which satisfies
Eq.~\eqref{eq:divErho} by construction. While this is a more straightforward approach, it
does not allow one to control the charge on each black hole.}
 $\phi$, giving
\begin{equation}
	\label{eq:phi_ansatz}
	E^i = \left(E_{\text{sp}}\right)^i + \nabla^i \phi.
\end{equation}
We then insert Eq.~(\ref{eq:phi_ansatz}) in Eq.~(\ref{eq:divErho}) with $\rho_\text{EM} = 0$ (since we assume the charge to be contained inside our excision surfaces) to get an elliptic equation for $\phi$
\begin{equation}
	\label{eq:phi}
	\nabla_j \nabla^j \phi = -\nabla_i (E_{\text{sp}})^i. 
\end{equation}
To solve Eq.~\eqref{eq:divB} for the magnetic field, we use an ansatz that satisfies the magnetic divergence constraint by construction, as described below.

\subsubsection{Choice of freely specifiable variables}
We are free to choose the background field $(E_{\text{sp}})^i$ in Eq.~\eqref{eq:phi_ansatz}. We set it to be the weighted superposition of the electric field of the two individual black holes.
\begin{equation}
	\label{eq:superposeE}
	(E_\text{sp})^i = \sum_{A=1}^2 e^{-r_{A}^2 / w_{A}^2} E^i_A,
\end{equation}
and solve for $\phi$ to construct
$E^i$ using Eq.~\eqref{eq:phi_ansatz}. See Appendix~\ref{append:KS} for expressions for the EM fields of a Kerr-Newman black hole in Kerr-Schild coordinates.

To find a magnetic field configuration that satisfies the constraints, we superpose the magnetic vector potentials
$\mathcal{A}^i_A$ of the two black holes in the same manner as we superpose the electric fields
\begin{equation}
	\label{eq:superposeA}
	(\mathcal{A}_{\text{sp}})^i = \sum_{A=1}^2 e^{-r_{A}^2 / w_{A}^2} \mathcal{A}_{A}^i, 
\end{equation}
and compute the magnetic field as in~\cite{Alcubierre:2009ij}, giving
\begin{equation}
	\label{eq:curlA}
	B^i = \dfrac{1}{\sqrt{\gamma}}\,\epsilon_F^{ikl} \partial_k (\mathcal{A}_{\text{sp}})_l.
\end{equation}
Since the divergence of a curl is zero, Eq.~\eqref{eq:curlA} automatically satisfies Eq.~\eqref{eq:divB}. Thus, we do not explicitly solve for the magnetic field, though the magnetic field of the final solution is affected by the rest of the solution through the contribution of the conformal factor to $\gamma$.

\subsubsection{Boundary conditions}
We want the potential $\phi$ to approach a constant at infinity for our isolated binary, so at the outer boundary we set
\begin{equation}
	\label{eq:bc_phi_inf}
	\phi = \text{const}\quad\text{at}~i_{0}.
\end{equation}
The specification of this constant is a gauge choice and we choose it to be zero. On the inner boundaries, we impose a Neumann boundary condition to control the charge of the black hole. Specifically, we scale the superposed electric field $E_{\text{sp}}$ on $\mathcal{S}$ to obtain the desired charge on each black hole. We set
\begin{equation}
	\label{eq:bc_phi_inner}
	s^i\partial_i\phi = \left(\dfrac{Q_{\text{d}, A}}{Q_{\text{sp},A}} - 1\right)\, s^i(E_{\text{sp}})_i\quad\text{on}~\mathcal{S}_{A}.
\end{equation}
In the above equation, $Q_{\text{d}, A}$ is the desired charge on the black hole (i.e., the same charge used to compute the background Kerr-Newman metric for that hole), and $Q_{\text{sp},A}$ is the charge on each black hole computed using the superposed electric field on each excision surface, i.e., 
\begin{equation}
	\label{eq:computeQ}
	Q_{\text{sp},A} =\dfrac{1}{4\pi}\oint_{\mathcal{S}_A} (E_{\text{sp}})^i \sqrt{h}\, dS_i,
\end{equation}
where $h$ is the determinant of $h_{ij}$ [which is given in Eq.~\eqref{eq:hij}], and $dS_i$ corresponds to the directed surface area element on the excision surface. To compute the charge for a black hole with the final solved electric field $E^i$, we replace  $(E_\text{sp})^i$ with $E^i$ in Eq.~\eqref{eq:computeQ}.

This construction will not work if the desired charge is nonzero and the superposed field leads to
zero charge (or a much smaller charge than the desired one). However, this sort of situation seems unlikely to occur in
practice, and we have not encountered it in our numerical investigations. Nevertheless, this is a relatively simple
construction to fix the charge. There is likely a better way to set boundary conditions on the electric field that is more in line with the isolated
horizon boundary conditions used for the geometry. We leave investigating such conditions for future work. 

In particular, the obvious requirement on the electric and magnetic fields at the horizon from the requirement that the excision surface be an isolated horizon (in fact, just a non-expanding horizon)
does not translate into a condition on the normal derivative of $\phi$. Specifically (see, e.g.,~\cite{Ashtekar:2000hw,Ashtekar:2004cn}) the relation
\begin{equation}
	\label{eq:ihEB}
	{}^{(4)}R_{\mu\nu} l^\mu p^\nu = 0
\end{equation} 	
is satisfied on the apparent horizon for any $p^\nu$ tangent to the
horizon. Here ${}^{(4)}R_{\mu\nu}$ is the $4$-dimensional Ricci
tensor, and $l^\mu$ is the outward-pointing null normal to the horizon. In our Einstein-Maxwell case, Eq.~\eqref{eq:ihEB} implies $F_{\mu\nu} l^\mu p^\nu = 0$ (see Sec.~II~C in~\cite{Ashtekar:2000hw}). Thus, from the expression for $F_{\mu\nu}$ in terms of the electric and magnetic fields (see Sec.~II in~\cite{Alcubierre:2009ij}), we have
\begin{equation}
\label{eq:ihEB_expanded}
[E_i^\parallel + (\mathbf{s} \times \mathbf{B})_i]\, p^i = 0.
\end{equation}
Here the cross product is the curved-space one from Eq.~\eqref{eq:cross_product_curved_space} and $E_i^\parallel := E_i - s_i s^j E_j$ is the projection of $E_i$ perpendicular to the
unit normal to the horizon, $s^i$.
Thus, Eq.~\eqref{eq:ihEB} implies that
\begin{equation}
	\label{eq:NEHrelation}
	E_i^\parallel = -(\mathbf{s} \times \mathbf{B})_i\quad\text{on}~\mathcal{S}_{A}.
\end{equation}
In Sec.~\ref{sec:results}, we discuss how well this condition is satisfied with our construction.

\section{Numerical Method}
\label{sec:num}
To construct initial data, we solve Eqs.~\eqref{eq:xcts} and~\eqref{eq:phi}, together with the boundary
conditions~\eqref{eq:bc_B_inf}, \eqref{eq:bc_shift_inner}, \eqref{eq:bc_psi_alphapsi_inner}, and \eqref{eq:bc_phi_inner}
using the pseudospectral code SGRID~\cite{Dietrich:2015pxa,Tichy:2009yr}. We use a Newton-Raphson scheme together with an
iterative generalized minimal residual solver with a block Jacobi preconditioner to solve the linearized equations during
each Newton step (see Chap.~4 in~\cite{Reifenberger:2013th} for details of the implementation). In the following sections,
we first describe our numerical grid and then discuss how we compute the ADM mass and linear and angular momentum of the
initial data as well as the quasilocal mass and spin of the black holes in our initial data solver.

\subsection{Surface Fitting Coordinates}
\label{sec:grid}
\begin{figure}[t]
\includegraphics[width=\columnwidth]{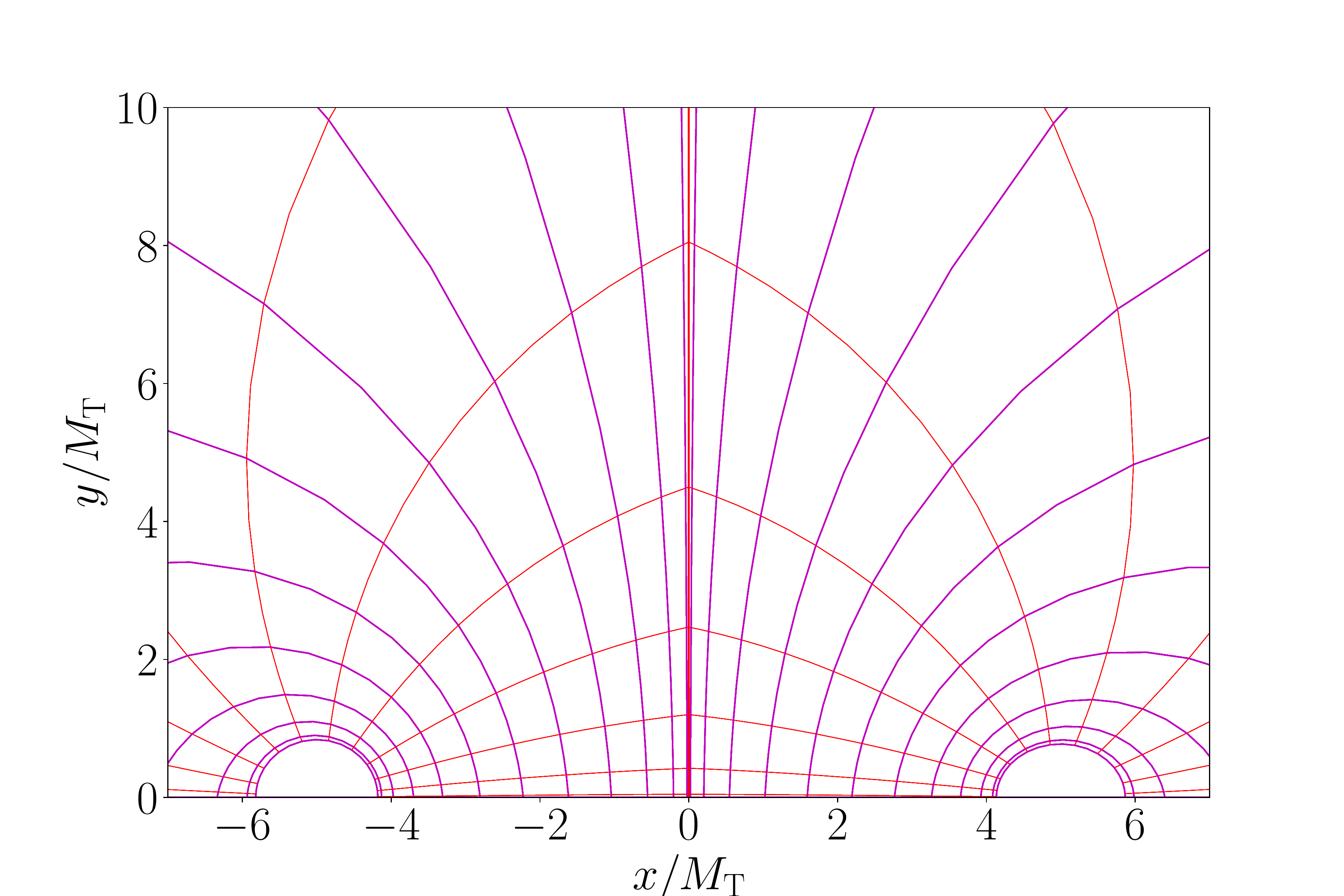}
\caption{\label{fig:grid}The $\phi = 0$ slice in the coordinates we use for the binary black hole configuration
\texttt{qc-sp7cp5} (see Table~\ref{tab:summary1}). The black holes are centered at $x = \pm5 M_{\text{T}}$, where $M_{\text{T}}$ is the total mass of the superposed black holes. Lines of constant $A_{\text{Ans}}$ and $B_{\text{Ans}}$ are shown using purple and red, respectively. The excision surfaces are
ellipsoids located at $A_{\text{Ans}} = 0$. The grid is separated into two computational subdomains that meet along the
$yz$-plane (seen here only as the line $x = 0$).}
\end{figure}

We use the compactified $(A_{\text{Ans}}, B_{\text{Ans}}, \varphi)$ coordinates introduced by Ansorg~\cite{Ansorg:2006gd}
to cover all of space outside the two excision regions with two computational domains (see Fig.~\ref{fig:grid}). The
coordinates $A_{\text{Ans}}$ and $B_{\text{Ans}}$ take values between 0 and 1, and $\varphi$ corresponds to the azimuthal
angle around the $x$-axis. $A_{\text{Ans}}= 0$ corresponds to the excision surface $\mathcal{S}_A$ in each computational
domain, and spatial infinity $i_0$ corresponds to the point $(A_{\text{Ans}}=1, B_{\text{Ans}} = 0)$ on the grid; see,
e.g., Chapter~5.2 of~\cite{Reifenberger:2013th}. At the interface between the two domains and on the $x$-axis, we impose
regularity conditions; see~\cite{Tichy:2009yr, Reifenberger:2013th}. For the spectral expansions, we use a Fourier basis
in the $\phi$ coordinate and Chebyshev bases in both $A_{\text{Ans}}$ and $B_{\text{Ans}}$ coordinates.

As in~\cite{Lovelace:2008tw}, we impose boundary conditions on the excision surface such that the excision surface
coincides with the apparent horizon of each black hole in the solved binary black hole initial data. We thus set the
excision surface around each black hole such that it coincides with the event horizon of the boosted Kerr-Newman black
hole used in the superposed metric. We do this by numerically solving for a freely specifiable function
$\sigma_A(B_{\text{Ans}},\varphi)$ that appears in the coordinate transformation between Cartesian coordinates $(x,y,z)$
and Ansorg coordinates $(A_{\text{Ans}},B_{\text{Ans}},\varphi)$. Specifically, we solve for $\sigma_{A}$ such that the
horizon equation
\begin{equation}
	\dfrac{\|\,\mathbf{r}_{\mathcal{S}} - (\hat{\boldsymbol{\chi}}\cdot \mathbf{r}_{\mathcal{S}})\,\hat{\boldsymbol{\chi}}\,\|^2}{r_+^2 + M^2\chi^2} + 	\dfrac{(\hat{\boldsymbol{\chi}}\cdot\mathbf{r}_{\mathcal{S}})^2}{r_+^2} = 1,
\end{equation}
is satisfied on each excision surface. Here, $\mathbf{r}_{\mathcal{S}}$ is the coordinate vector pointing from the black
hole center to a point on the excision surface (located at $A_{\text{Ans}}=0$). For a black hole hole with nonzero
velocity, we apply the appropriate Lorentz transformation to $\mathbf{r}_{\mathcal{S}}$ to account for the length
contraction of the horizon due to the boost. Further, $M$, $\chi$, and $\hat{\boldsymbol{\chi}}$ represent the black hole's mass, the
magnitude of its dimensionless spin, and the unit vector along the spin axis, respectively. Additionally, $r_{+}$ is the radius
of the outer horizon of a Kerr-Newman black hole given by
\begin{equation}\label{eq:r_plus}
	r_{+} = M + M\sqrt{1 - \chi^2 - \mathcal{Q}^2},
\end{equation} 
where $\mathcal{Q} := Q / M$, and $Q$ is the charge of the Kerr-Newman black hole.

\subsection{Computing diagnostics}
\label{subsec:ADM}
In order to characterize our initial datasets and control our initial data parameters, we compute the ADM mass and linear and angular momenta of the initial data. We also compute
quasilocal measures of the mass and spin of each individual black hole.

To compute the ADM mass, we follow~\cite{Marronetti:2007ya} in obtaining a more numerically accurate expression by writing the original surface integral at infinity as the sum of a volume integral and surface integrals over the excision surfaces and removing the second derivatives of the conformal factor using the Hamiltonian constraint. We correct the expressions from that paper for a flipped pair of indices in the integrand of the surface integral and generalize to the conformally curved case; cf.\ the original versions of the expressions in~\cite{Yo:2002bm}. In addition, we also incorporate the source terms arising from the Hamiltonian constraint to obtain 
\begin{widetext}
\begin{equation}
\begin{split}
    \label{eq:MADM}
    M_{\rm ADM} &= \dfrac{1}{16 \pi} \int_{\mathcal{V}} \biggl[(1-\psi)~\tilde{R} + \tilde{\Gamma}^k ~\tilde{\Gamma}^{i}_{~ki} - \tilde{\Gamma}^{ikj} \tilde{\Gamma}_{kij} 
                + \psi^{-7}\tilde{A}^{jk}\tilde{A}_{jk} + \psi^{5}\left(16 \pi \rho  - \dfrac{2}{3} K^2\right) \biggr] \sqrt{\tilde{\gamma}}  ~dV \\
                  &\quad + \dfrac{1}{16 \pi} \sum_{A=1}^2 \oint_{\mathcal{S}_A} \psi^4 (\tilde{\Gamma}^k - \tilde{\Gamma}^{ik}_{~~~i} - 8~ \tilde{\nabla}^k \psi) \sqrt{\tilde{h}}~dS_k,
\end{split}
\end{equation}
\end{widetext}
where $\tilde{R}$ and $\tilde{\Gamma}^i_{jk}$ are the Ricci scalar and Christoffel symbols computed using the conformal metric and $\tilde{\Gamma}^i := \tilde{\gamma}^{jl} \tilde{\Gamma}^i{}_{jl}$. Additionally, $\mathcal{V}$ is the region outside the excision surfaces and $dV$ is the volume element in flat space.

To compute the ADM linear and angular momentum, we follow~\cite{Ossokine:2015yla}, adding the relevant source terms due to the presence of the EM fields to get
\begin{subequations}
	\label{eq:ADM}
\begin{align} 
    \label{eq:PADM}
    P^i_{\rm ADM} &= \frac{1}{8\pi}\sum_{A=1}^2\oint_{\mathcal{S}_A} P^{ij}\,dS_j - \frac{1}{8\pi} \int_\mathcal{V} G^i\,dV, \\
    \label{eq:JADM}
    J^z_{\rm ADM} &= \frac{1}{8\pi}\sum_{A=1}^2\oint_{\mathcal{S}_A} (x P^{yj}-yP^{xj})\,dS_j \\\nonumber 
                    &\quad - \frac{1}{8\pi} \int_{\mathcal{V}}(xG^y-yG^x)\,dV,
\end{align}
\end{subequations}
where $J^x_{\rm ADM}$ and $J^y_{\rm ADM}$ are obtained by cyclic permutations of $(x,y,z)$ in the above equation, and
\begin{subequations}
\begin{align}
    \label{eq:Pij}
    P^{ij} &:= \psi^{10}(K^{ij}-K\psi^{-4}\tilde{\gamma}^{ij}),\\
    \label{eq:Gij}
    G^i &:= \tilde\Gamma^i_{jk}P^{jk} +\tilde\Gamma^j_{jk}P^{ik} - 2\tilde{\gamma}_{jk}P^{jk} \tilde{\gamma}^{il}\partial_l(\ln\psi) - 8 \pi J^i.
\end{align}
\end{subequations}
As in~\cite{Ossokine:2015yla}, we apply a rolloff to the volume integrands at large radii in Eqs.~\eqref{eq:ADM}
to reduce the contributions from numerical noise in the regions near infinity where the integrand is
small and the volume element is large. Specifically we set
\begin{equation}
G^i_{\text{rolloff}} =
  \begin{cases}
  	G^i & \text{if $r \leq R_c$} \\
  	(R_c^2 / r^2)~ G^i & \text{if $r > R_c$},
  \end{cases}
\end{equation}
where $r$ is the coordinate distance from the origin, and $R_c$ is the roll-off radius. We apply the same roll-off
to the volume integrand in Eq.~\eqref{eq:MADM}. We set $R_c = 500M_\text{T}$,
where $M_\text{T}$ is the sum of the masses of the two superposed Kerr-Newman black holes.

We also compute the quasilocal mass and spin of the black holes in the standard way, through integrals over the excision surfaces, with just a small generalization to the charged case.
Specifically, we compute the irreducible mass of the horizon
\begin{equation}
M_\text{irr} = \sqrt{\frac{\mathcal{H}}{16\pi}},
\end{equation}
where $\mathcal{H}$ is the area of the horizon. We also compute the angular momentum of the
horizon $\mathcal{J}^k$ using the standard isolated horizon integral~\cite{Schnetter:2006yt} with the contribution from the EM fields as
in~\cite{Ashtekar:2001is}, giving
\begin{equation}
	\label{eq:IHangularmomentum}
	\mathcal{J}^k = \frac{1}{8\pi}\oint_{\mathcal{S}_{\text{A}}}[K_{ij} + 2 (\mathcal{A}_{\rm sp})_i E_j ]\,\xi^{i}_{~(k)} \sqrt{h}\, dS^j,
\end{equation}
where $(\mathcal{A}_{\rm sp})_i$ is given by Eq.~\eqref{eq:superposeA} and $\xi^{i}_{~(k)}$ are the flat space Killing
vectors defined in Eq.~\eqref{eq:Killingvec}. The sign of the EM term is opposite the one given in~\cite{Ashtekar:2001is},
since we found that this gives the correct result for an isolated Kerr-Newman black hole. We presume that this is due to a
difference in sign convention, particularly since~\cite{Bozzola:2019aaw,Bozzola:2021elc} use the
same sign as~\cite{Ashtekar:2001is}, but have not been able to find the exact source of this difference. We then compute the horizon mass, often known as the Christodoulou-Ruffini mass,
$M_{\text{Chr}}$~\cite{Christodoulou:1971pcn}, using
\begin{equation}
	M_{\text{Chr}}^2 = \left(M_{\text{irr}} + \frac{Q^2}{4 M_{\text{irr}}}\right)^2 + \frac{\mathcal{J}^2}{4M_{\text{irr}}^2},
\end{equation}
where $\mathcal{J}$ the magnitude of the angular momentum given by Eq.~\eqref{eq:IHangularmomentum}, and $Q$ is the horizon charge given by Eq.~\eqref{eq:computeQ} (computed using the solved electric field). In Sec.~\ref{sec:results}, we use the sum of the Christodoulou masses, denoted $M_{\rm{T}, \rm{C}}$, which gives the total mass of the binary at infinity. 

\subsection{Controlling BH spin and ADM linear momentum}
\label{subsec:iter}
In Sec.~\ref{sec:xctsbc}, we introduced two parameters in the boundary conditions, $v_0^i$ in Eq.~\eqref{eq:bc_B_inf} and
$\Omega_r^i$ in Eq.~\eqref{eq:bc_shift_inner}, which we use to control the ADM linear momentum and the black hole spins,
respectively. We discuss how we set them here. As in~\cite{Ossokine:2015yla}, we iteratively set $v_0^i$ in
Eq.~(\ref{eq:bc_B_inf}) to drive the ADM linear momentum to zero. However, we use a simpler iterative procedure, setting
\begin{equation}
	\label{eq:veliter}
v_{0,n+1}^i = v_{0,n}^i - \frac{P_{\text{ADM},n}^i}{M_\text{T}},
\end{equation}
where $n$ indexes the Newton iterations. We start the iteration from $v_{0,0}^i = 0$.

We similarly drive the dimensionless spins to their desired values (those of the superposed Kerr-Newman metrics) by
adjusting $\Omega_r^i$ using a simple iteration inspired by the form of the Kerr-Newman horizon angular
velocity~\cite{Newman:2014}. We start with the angular velocity of the Kerr-Newman spacetime used in the superposition
\begin{equation}
	\Omega_{r,0}^i = \frac{\mathcal{J}^i_\text{KN}}{4M_\text{Chr, KN} M_\text{irr, KN}^2},
\end{equation}
where $M_\text{irr, KN}$, $M_\text{Chr, KN}$, and $\mathcal{J}^i_\text{KN}$ are the irreducible mass, horizon mass (i.e., Kerr-Newman mass parameter), and angular momentum of each superposed black hole, respectively. We then iteratively update $\Omega_{r}^i$ using
\begin{equation}
	\Omega_{r, n+1}^i = \Omega_{r,n}^i + \frac{M_{\text{Chr},n}^2\chi^i_\text{KN} - \mathcal{J}_{n}^i}{4M_{\text{Chr},n}M_{\text{irr},n}^2},
\end{equation}
where, as in Eq.~\eqref{eq:veliter}, $n$ indexes the Newton iterations. Thus, e.g., $\mathcal{J}_n^i$ is the horizon
angular momentum recomputed after the $n^\text{th}$ iteration on the excision surface under consideration. Additionally,
$\chi^i_\text{KN} := \mathcal{J}_\text{KN}^i/M_\text{Chr, KN}^2$ is the dimensionless spin vector of the corresponding
Kerr-Newman metric used in the superposition.

\section{Results}
\label{sec:results}
We now discuss the initial data we constructed to test the code, particularly its convergence with resolution, and the exploratory evolutions we performed. We summarize the cases we consider in Tables~\ref{tab:summary1} and~\ref{tab:summary2}. 

\begin{table*}[t]
\begin{center}
\begin{tabular}{ccccccccccc}
\toprule
{Name} & {$q$} & {$\mathcal{Q}_1$} & {$\mathcal{Q}_2$} & {$\boldsymbol{\chi}_1$} & {$\boldsymbol{\chi}_2$} & {$M_\text{T}\Omega_0$} & {$d/M_{\text{T}}$} & {$\dot{a}_0$} & {$\zeta$} &{Evol} \\
\hline
 {\texttt{qc-sp5}} & {1} & {0} & {$0$} & {(0, 0, 0.50)} & {(0, 0, 0.50)} & {0.01941} & {10} & $\hphantom{-}0.006531$ & 0.5 &{BAM} \\
 {\texttt{qc-hs}} & {1.16} & {0} & {$0$} & {(0, 0, 0.69)} & {(0, 0, 0.79)} & {0.02848} & {16} & {$0$} & 1.25 & -- \\
 {\texttt{qc-hc}} & {1.16} & {0.97} & {$-0.97$} & {(0, 0, 0)} & {(0, 0, 0)} & {0.02848} & {10}  & {$0$} & 2 & -- \\
 {\texttt{qc-mc}} & {1.16} & {0.59} & {$-0.45$} & {(0, 0, 0)} & {(0, 0, 0)} & {0.02848} & {10}  & {$0$} & 2 & -- \\
 {\texttt{qc-sp7cp5}} & {1.16} & {0.56} & {$-0.43$} & {(0, 0.69, 0)} & {(0.47, 0, 0)} & {0.02848} & {10} & {$0$} & 2 &-- \\
\toprule
\end{tabular}
\caption{\label{tab:summary1} Summary of the parameters of the initial data sets for orbiting binaries (``qc'' for
``quasicircular''), also indicating the case for which we performed a test evolution with BAM. Here $\mathcal{Q}_{A}$ and
$\boldsymbol{\chi}_{A}$ are the dimensionless charge and spin on black hole $A$, respectively. Additionally, $q$, $\Omega_{0}$,
and $\dot{a}_0$ correspond to the mass ratio, angular velocity, and radial velocity of the binary, respectively. Finally,
$d$ is the separation between the two black holes and $\zeta$ is the attenuation width parameter defined in
Eq.~\eqref{eq:zeta}. We set the velocity of each black hole using the Newtonian expression in terms of $\Omega_0$ and $d$.
}
\end{center}
\end{table*}

\begin{table}
\begin{center}
\begin{tabular}{cccc} 
\toprule
 {Name} & {$\mathcal{Q}_1 = \mathcal{Q}_2$}  & {$v$} & {Evol} \\
 \hline
 {\texttt{ho-v0q0}}  & {0.0}  & {0.0} & {HAD, BAM} \\
 {\texttt{ho-v0qp1}} & {0.1} & {0.0} & {HAD} \\
 {\texttt{ho-v0qp3}}  &  {0.3} & {0.0} & {HAD} \\
 {\texttt{ho-v0qp5}} & {0.5} & {0.0}  & {HAD} \\ 
 {\texttt{ho-vp1q0}} & {0.0} & {0.1} & {HAD} \\
 {\texttt{ho-vp3q0}} & {0.0} & {0.3} & {HAD} \\
\toprule
\end{tabular}
\caption{\label{tab:summary2} Summary of the parameters of the initial data sets for head-on collisions (abbreviation ``ho'' ) of equal-mass, equal-charge, nonspinning binaries we constructed and performed test evolutions of with HAD and BAM. Here, $v$ is the magnitude of the velocity of each black hole. We set $d=10M_{\text{T}}$ and $\zeta=2$ for all configurations listed here and set $\dot{a}_0 = -v/d$ for the head-on collisions.}
\end{center}
\end{table}

\subsection{Initial Data}

\begin{figure}
\includegraphics[width=\columnwidth]{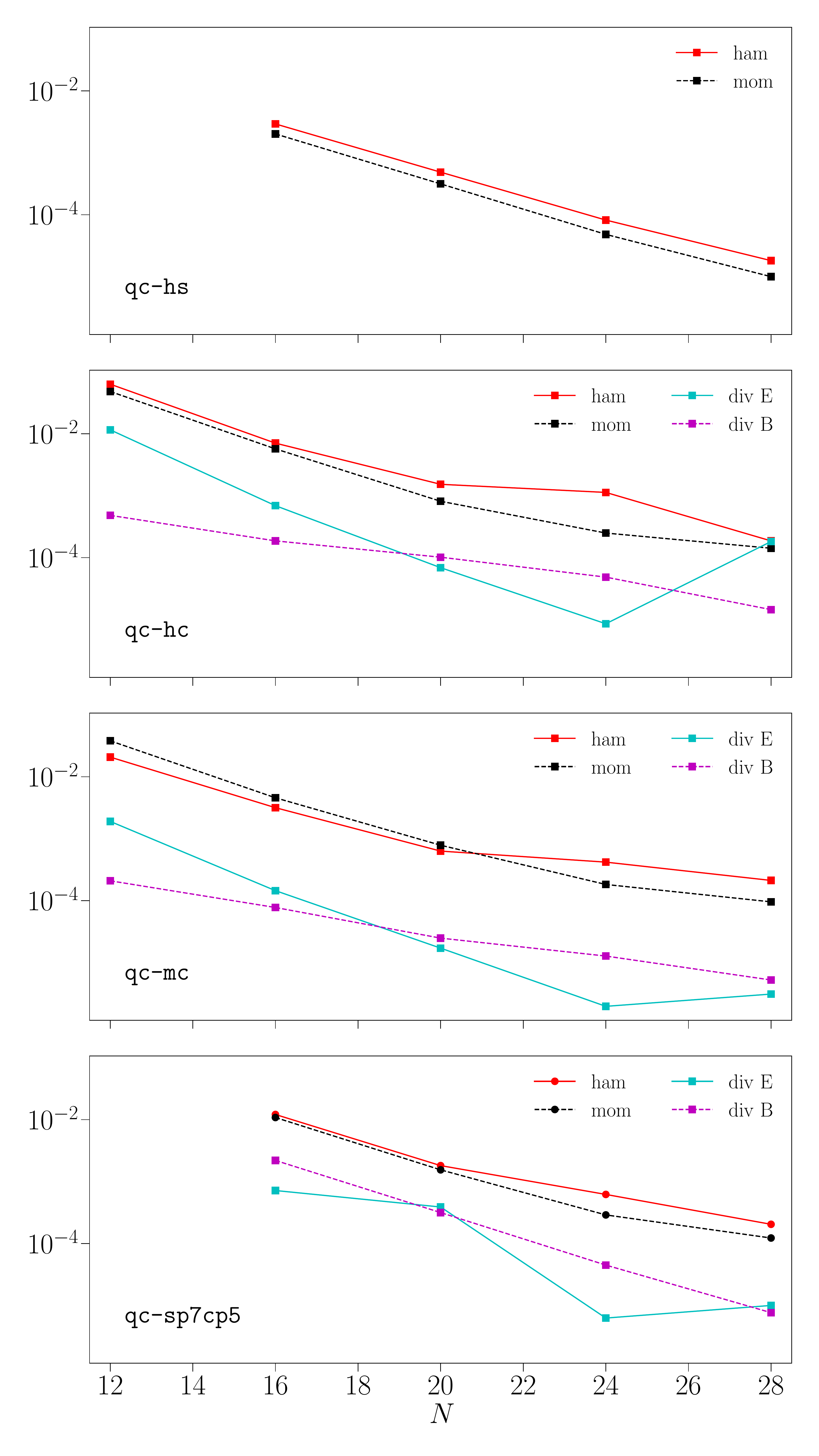}
\caption{\label{fig:convergence_high_charge} Convergence of the constraint residuals for four different binary configurations,
a highly spinning case (\texttt{qc-hs}), highly charged (\texttt{qc-hc}) and moderately charged (\texttt{qc-mc}) cases,
and a more generic charged and spinning case (\texttt{qc-sp7cp5}). In all cases, we show the
Hamiltonian and momentum constraints (``ham'' and ``mom''), and for the charged cases, we also show the electric
and magnetic constraints (``div E'' and ``div B''). The horizontal axis gives the number of points used for each
coordinate ($N$) and the vertical axis shows the $L^2$ norm of the physical constraints
over the entire computational grid (scaled by the total volume of the computational grid).
Hence, we scale the $L^2$ norm by the square of the total Christodoulou mass $M_{\rm{T},\rm{C}}$.
For the momentum
constraint, this includes the Euclidean vector norm on the components of the constraint.
For $N=12$, the initial data solver does not converge for \texttt{qc-hs} and
\texttt{qc-sp7cp5}, since the larger gradients near the horizon from significant spin require more resolution to reach the
convergent regime.} 
\end{figure}

To test the convergence of our initial data solver, we consider four representative
orbiting configurations: two nonspinning cases with high (\texttt{qc-hc}) and moderate (\texttt{qc-mc})
charge, one uncharged case with reasonably high spins (\texttt{qc-hs}), and finally, a more
generic precessing binary with moderate charges and spins (\texttt{qc-sp7cp5}). We choose opposite signs for the charges in the
charged cases to increase the asymmetry and thus provide a more stringent test of the solver. We use
the same number of points ($N$) in the $A_{\text{Ans}}, B_{\text{Ans}}$, and $\phi$
directions for each configuration, and set
\begin{equation}
	\label{eq:zeta}
	w_A = \zeta \dfrac{M_A}{M_1 + M_2} d, 
\end{equation}
for the attenuation weights in Eqs.~\eqref{eq:sks}, \eqref{eq:superposeE}, and~\eqref{eq:superposeA}. Here, $\zeta$ is a
dimensionless free parameter and $d$ is the coordinate distance between centers of the two black holes. We set $\Omega_0$
and $\dot{a}$ in Eq.~\eqref{eq:bc_shift_inner} by performing iterative eccentricity reduction. For
the other initial data sets, we set $\Omega_0$ using the PN expression in~\cite{Mroue:2012kv} for \texttt{qc-sp5}, and set $\dot{a} = 0$. As we discuss later, we do
not expect these settings to give low-eccentricity initial data, so we do not attempt to include the effects of the charge. In
Fig.~\ref{fig:convergence_high_charge} we show the convergence of the geometric (Hamiltonian and momentum) and EM
constraint residuals for all the four cases. We find the expected exponential (spectral) convergence at low resolutions,
but find slower convergence at higher resolution.

In particular, for the geometric constraints, we see spectral convergence only up to $N=20$ for the highly charged case
(\texttt{qc-hc}). For the highly spinning case (\texttt{qc-hs}), the convergence is exponential up to $N=28$. For the
binary with more generic parameters (\texttt{qc-sp7cp5}), the rate of convergence is better than for the highly charged
case (possibly due to the fact that the more generic system is less extreme), but the Hamiltonian constraint shows
subexponential decay after $N = 20$. While this looks similar to the subexponential convergence found for the construction
without attenuation functions in~\cite{Lovelace:2008hd}, the cause of the slower than expected convergence in our case
must be due to a different cause: We use attenuation functions when superposing both the metric and the electromagnetic
fields, which ensures that they fall off rapidly at infinity, so there is no possibility for the logarithmic term in the
solution that caused the subexponential convergence in~\cite{Lovelace:2008hd}. In fact, we do not even have the
co-rotation term in the outer boundary condition for the shift that leads to that logarithmic term in the Hamiltonian
constraint when combined with the $1/r$ falloff of the conformal metric without attenuation functions. One other possible
cause that we can exclude is the adjustments to $v_0$ and $\Omega^i_r$ that we perform during the Newton iterations. We
disabled these iterations but found no improvement in the speed of convergence. For the EM constraints, we see spectral
convergence for electromagnetic constraints up to $N=24$ for all the configurations except for \texttt{qc-sp7cp5} at
$N=20$. The rate of exponential decay, however, is much slower for the magnetic constraint than for the electric
constraint.

\begin{figure}
\includegraphics[width=\columnwidth]{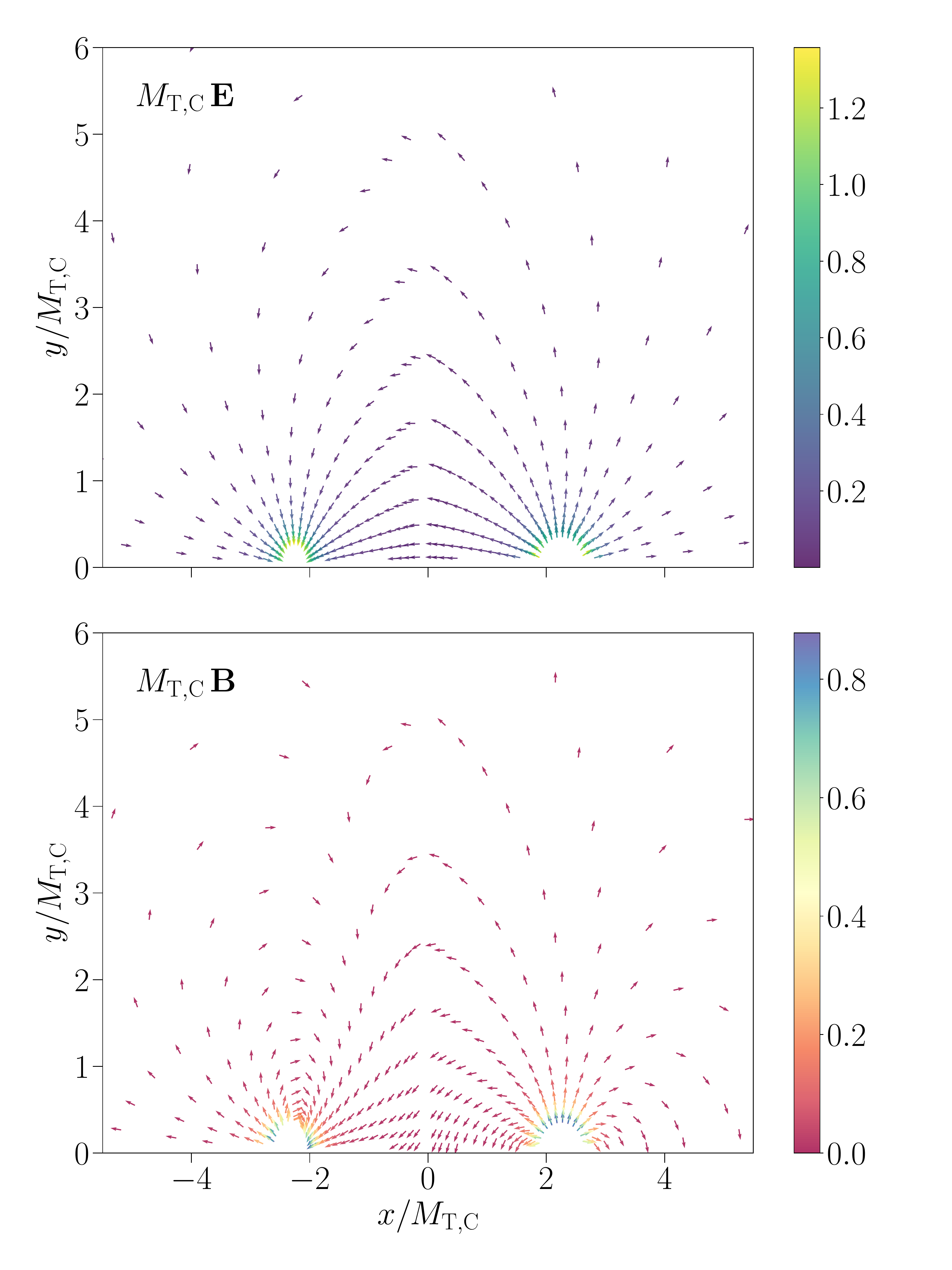}
\caption{\label{fig:EBfields} The electric (top panel) and magnetic (bottom panel) fields (multiplied by
$M_{\rm{T},\rm{C}}$) at the grid points around the two black holes in the $xy$-plane for \texttt{qc-sp7cp5}. The arrows
represent the unit vectors of the projection of the field into the $xy$-plane. The magnitude of the projection is shown by
the color scheme. The black hole on the left (right) has its spin aligned along the $x$- ($y$-) axis giving rise to two
magnetic dipole moments aligned with the two spin axes.}
\end{figure}

We also illustrate the solved electric and magnetic fields for the \texttt{qc-sp7cp5} case, with its oppositely charged,
spinning black holes, in Fig.~\ref{fig:EBfields}. As one would expect, we observe an overall electric dipole moment
aligned along the axis joining the two black holes, and a magnetic dipole for each black hole aligned with its spin axis.
Further, the magnetic field due to the orbital motion of the black holes (not visible in the figure, since it is
perpendicular to the plane plotted and thus projected out) is consistent with the magnetic field around two charges with
different signs moving in opposite directions. Unsurprisingly, we find the largest corrections to the unsolved superposed
electric and magnetic fields near each black hole.

Finally, we check how well the isolated horizon condition on the EM fields at the horizon [Eq.~\eqref{eq:ihEB_expanded}]
is satisfied in our initial data. We do this by computing the $L^2$ norm of the residual over the horizon, as well as the
version with the sign reversed, for comparison. Specifically, we define
\begin{subequations}
\begin{align}
	Z^{\pm}_i &:= E_i^\parallel \pm (s \times B)_i, \\
\text{EB}_{\text{res},\pm}^2 &:= \oint_{\mathcal{S}_\text{A}} \gamma^{ij} Z^{\pm}_i Z^{\pm}_j\, \sqrt{h}\, s^k dS_k,
\end{align}
\end{subequations}
where $\text{EB}_{\text{res},+}/\text{EB}_{\text{res},-}$ gives a dimensionless measure of how well the relation is
satisfied on each excision surface. We found the ratio to be the lowest (0.04) for the larger black hole in
\texttt{qc-sp7cp5} and the largest (0.09) for the smaller black hole in \texttt{qc-mc}. Additionally, in the nonspinning cases, we found this ratio
to depend on the asymmetry of the magnitude of the two dimensionless charges. Hence, even
though the black holes in $\texttt{qc-mc}$ have smaller charges than in $\texttt{qc-hc}$, we find the deviations from the
isolated horizon condition to be larger for $\texttt{qc-mc}$ which has the largest dimensionless charge ratio among
the charged, nonspinning configurations.

\subsection{Exploratory Evolutions}
\label{sec:evolutions}
We performed test evolutions of vacuum data initial data using the BAM code~\cite{Brugmann:2008zz}. For the charged case,
we used the HAD code~\cite{Liebling:2020jlq} to evolve head-on collisions of charged black holes. Both of these codes,
however, are designed to evolve puncture initial data, and hence require black hole filling to evolve our excision initial
data (see, e.g.,~\cite{Brown:2008sb}). We used BHfiller (described in Sec.~3.2 of~\cite{Reifenberger:2013th}) to fill
inside the excision surfaces in BAM. We had to generalize the filling algorithm slightly to account for our nonspherical
excision surfaces (see Appendix~\ref{append:BHFiller}). HAD uses a different approach to filling, described in
Appendix~\ref{append:BHFillerHAD}.

\subsubsection{Uncharged binary black holes in orbit}
For the uncharged case, we evolved an equal-mass, equal-aligned-spin quasicircular binary inspiral in orbit
(\texttt{qc-sp5}, with dimensionless spins of $0.5$ in the direction of the orbital angular momentum) using BAM with the
BSSN formulation of the equations and the standard puncture gauge. For this initial test, we used $16$~points in each
direction to construct the initial data using SGRID. For the evolution, we use seven refinement levels with three moving
levels and four fixed levels. Each refinement level has half the grid spacing of the previous one with a minimum grid
spacing of $0.0625M_\text{T}$. The outer boundary of the computational domain is at $\sim 250M_\text{T}$. We use
fourth-order spatial finite differencing and fourth-order integration in time, with a Courant factor of $0.25$. We extract
the gravitational waves at a radius of $r_{\text{ext}}=50M_\text{T}$.

We reduced the eccentricity by adjusting $\Omega_0$ and $\dot{a}$ using the iterative method in~\cite{Pfeiffer:2007yz},
where we measure the eccentricity using the puncture tracks. We found that the eccentricity one obtains from the
post-Newtonian values for $\Omega_0$ and $\dot{a}_0$ given in~\cite{Mroue:2012kv} is large enough that the iterative
method does not reduce the eccentricity when starting from those values. We thus adjusted those parameters by hand until
the eccentricity was small enough that the iterative method produced further reductions of the eccentricity. We obtained
an eccentricity of about $0.06$, which is relatively large, compared to the eccentricities needed for waveform modeling
(e.g., the eccentricities of $\sim 10^{-3}$ achieved for some of the simulations produced in~\cite{Husa:2015iqa}). While
we could have carried out further iterations of the eccentricity reduction procedure, we chose not to for this initial
test. In particular, the unusual nature of the eccentricity reduced setup we obtained, where the coordinate separation
between the punctures increases before decreasing when the data are evolved, deserves more careful investigation.

\begin{figure}[t]
\includegraphics[width=\columnwidth]{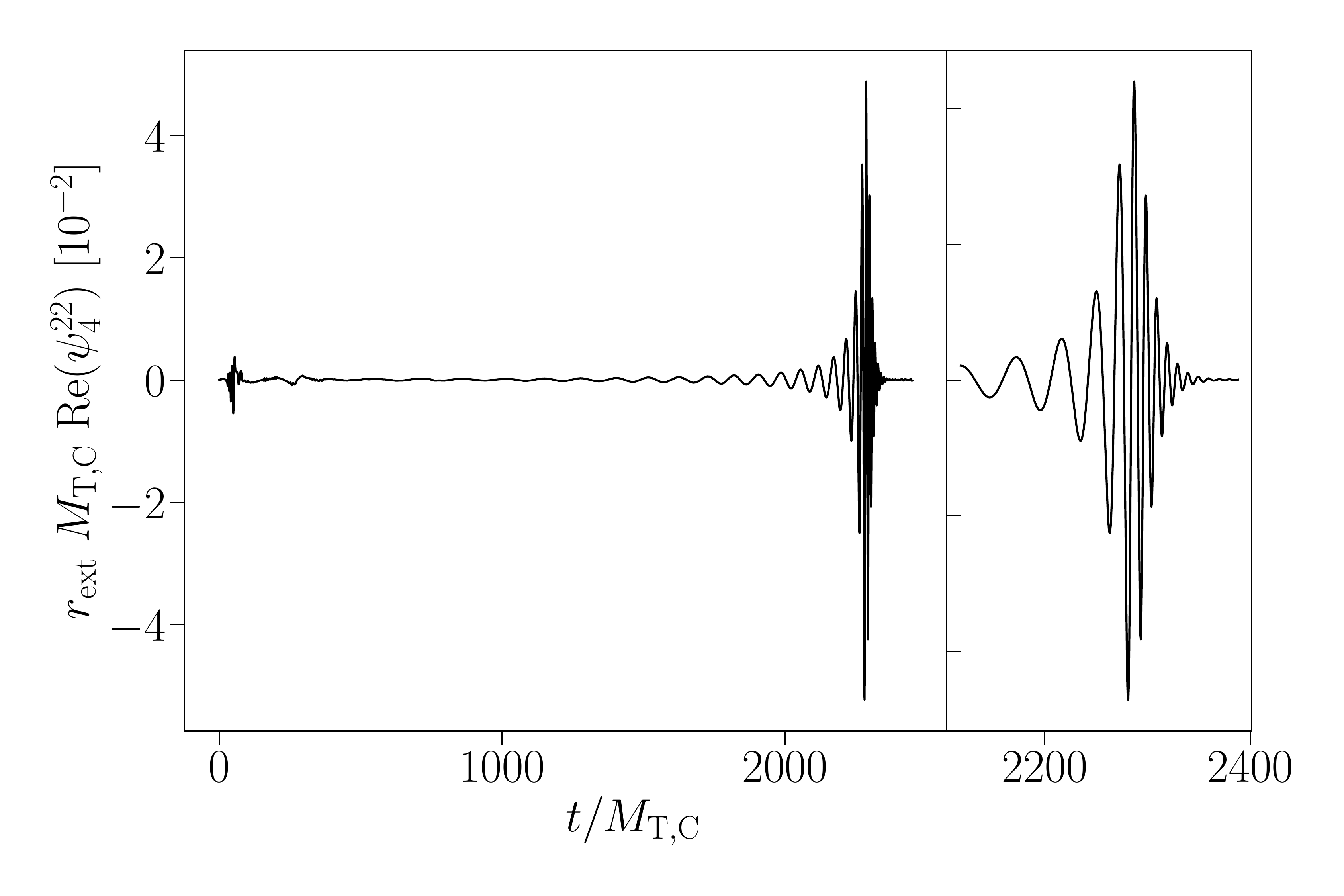}
\caption{\label{fig:BAMpsi4}The real part of the $l = m =2$ mode of the Weyl scalar $\psi_4$ from the uncharged,
quasicircular, aligned-spin evolution \texttt{qc-sp5}. The left panel shows the complete waveform, and the right panel
shows the waveform around the merger.}
\end{figure}

\begin{figure}[t]
\includegraphics[width=\columnwidth]{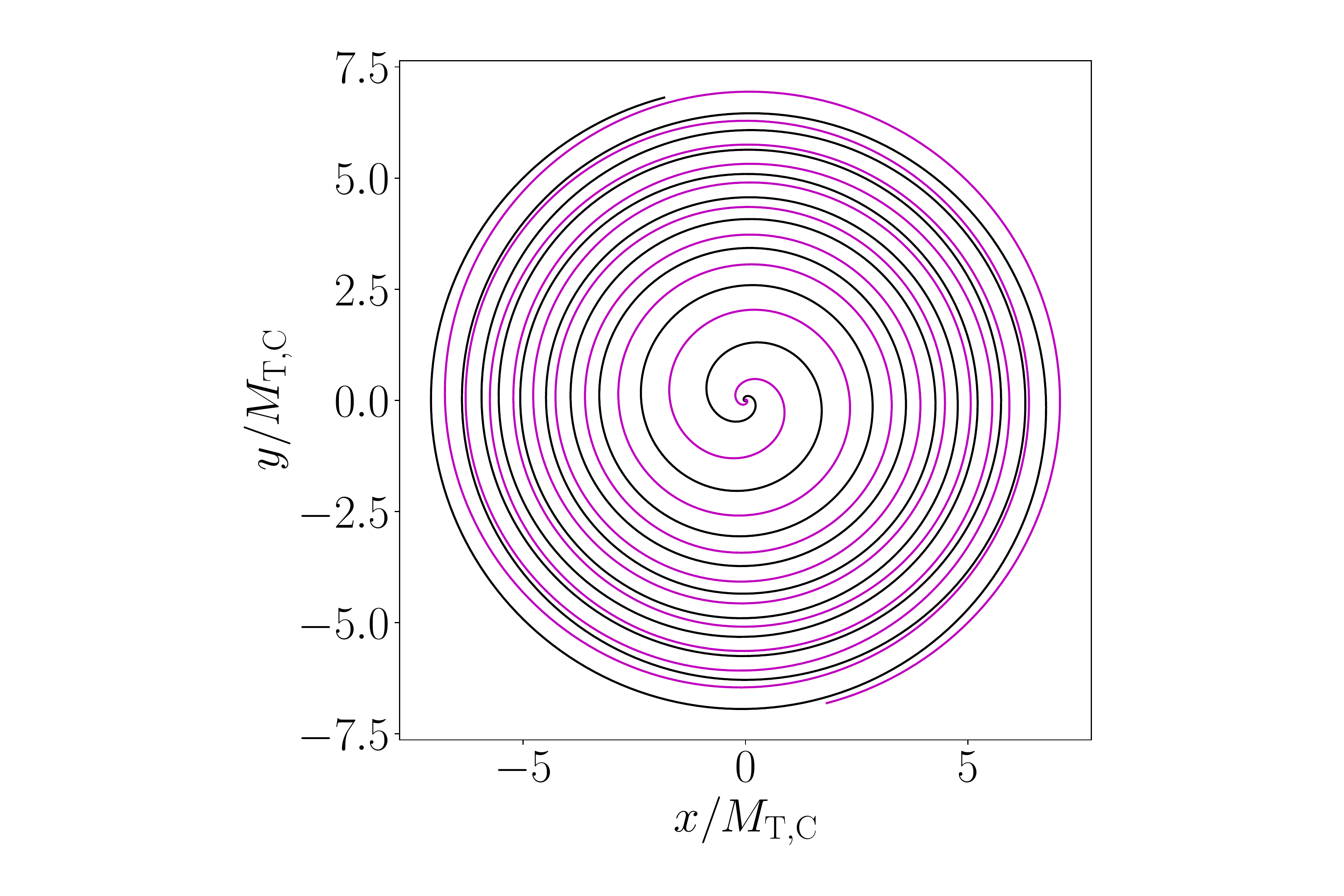}
\caption{\label{fig:BAMpuncturetracks}Puncture tracks of the two black holes for the \texttt{qc-sp5} evolution. The tracks
show approximately 8 orbits before merger and demonstrate the nonnegligible eccentricity of $\sim 0.06$.}
\end{figure}

In Figs.~\ref{fig:BAMpsi4} and~\ref{fig:BAMpuncturetracks}, we show the real part of the quadrupolar ($l=m=2$) mode of the
Weyl scalar $\psi_4$, and the puncture tracks for the last $\sim 8$ orbits before merger, respectively, for
\texttt{qc-sp5}, to illustrate that they are qualitatively as expected. As is clear from the nearly overlapping puncture
tracks, the binary still has a relatively large eccentricity (consistent with the value of $\sim 0.06$ obtained above).
An important consideration when constructing initial data is how much the properties of the binary (e.g., black hole spin)
change during early states of the evolution, as the system relaxes and emits junk radiation. We observed that the dimensionless spins settle to $0.49$ (compared to the desired value of $0.5$) after the first $900M_{\rm{T,C}}$ during which the system relaxes by emitting junk radiation. The relative error between the computed and desired spins decreases through the first $900M_{\rm{T,C}}$ and is $\sim6\%$ at its maximum. 

\subsubsection{Head-on collisions of charged binary black holes}
We now consider head-on collisions of charged black holes. Unfortunately, we were unable to produce evolutions of orbiting
binaries with HAD, even when evolving a vacuum case for which the BAM evolution orbited. The evolutions of orbiting
initial data with HAD initially resemble quasi-circular orbits, but quickly the trajectory of the black holes becomes
nearly head-on. This failure to orbit may be due to the method of filling the excised interior developed in HAD for this
project, described in Appendix~\ref{append:BHFillerHAD}. In particular, even though~\cite{Brown:2008sb} showed that the constraint violations from the filling do not
propagate outside the horizon in BSSN, the finite difference stencils extend inside the horizon and thus the filling
affects the evolution through the derivatives of the fields. In fact,~\cite{Brown:2008sb} finds that one needs the filling
to extend out to a coordinate radius at most $0.4$ times the horizon radius in order for the filling not to affect the
evolution (though the isotropic coordinate system of their initial data is rather different from our Kerr-Schild one).
However, for our initial data construction, the apparent horizon coincides with the excision surface, and so it is
necessary to fill all the way to the apparent horizon. It thus might be worthwhile to extend the construction to allow for
the excision surface to be inside the apparent horizon by using negative expansion boundary conditions, as
in~\cite{Varma:2018sqd}.

For our tests of charged, head-on collisions, we evolved equal-mass, equal-charge, nonspinning binaries, as well as some
uncharged, boosted head-on collisions (see Table~\ref{tab:summary2} for an overview of these cases). The charged cases
allow us to compare with previous numerical work~\cite{Zilhao:2012gp, Zilhao:2013nda}. In particular as discussed below,
analytical scalings found in their work agree well with our results.

We used a resolution of $(18, 18, 12)$ points in the $A_{\text{Ans}}, B_{\text{Ans}}$, and $\phi$ directions to generate
the initial data in SGRID. The HAD evolutions used four refinement levels and each refinement level has half the grid
spacing of the previous one, with a minimum grid spacing of $0.0891M_\text{T}$. The outer boundary of the
computational domain is at $\sim 72M_\text{T}$. We used fourth-order spatial finite differencing and third-order
integration in time with a Courant factor of $0.25$. We extract gravitational and electromagnetic waves at
$r_{\text{ext}}=50M_{\rm{T}}$. For the electromagnetic emission, we compute the scalar function $\Phi_2$ that contains
the transverse radiative degrees of freedom of the electric field, in the asymptotic limit~\cite{Newman:1961qr}. The
details of the evolution system (BSSN with the standard puncture gauge plus constraint damping for the EM equations) and
code are described in~\cite{Hirschmann:2017psw} which evolved binary black holes in Einstein-Maxwell-dilaton theory. For
these evolutions we simply set the dilaton coupling parameter $\alpha$ to zero.

\begin{figure}[t]
\includegraphics[width=\columnwidth]{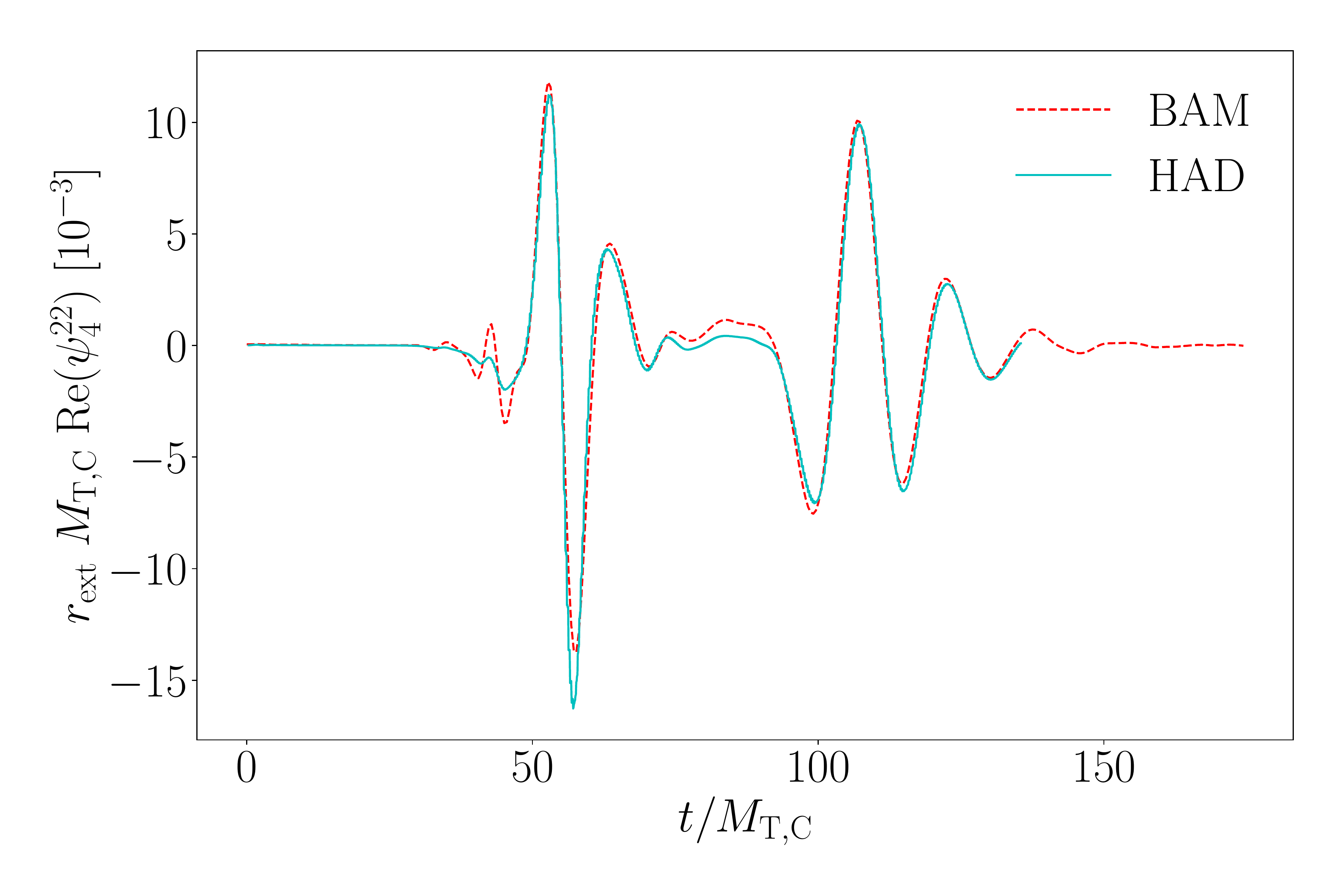}
\caption{\label{fig:HAD-BAM} A comparison of the real part of the $l = m =2$ mode of $\psi_4$ from the head-on collision of two uncharged black holes starting from rest (\texttt{ho-v0q0}) evolved using  HAD and BAM. The waveforms are not aligned in any way.}
\end{figure}

In Fig.~\ref{fig:HAD-BAM}, we compare the quadrupolar mode of $\psi_4$ between HAD and BAM evolutions for a head-on
collision of two uncharged black holes starting from rest. Both waveforms are extracted at $r_{\text{ext}} = 50
M_{\text{T}}$. Note that while the junk radiation profiles show some differences, the merger section of the waveform
is in good agreement. While we have not attempted to quantify the error budgets of the two simulations, we anticipate that
the differences seen are well within the combined error budget, particularly since these are not particularly high
resolution simulations and have rather small GW extraction radii.

\begin{figure}[t]
\includegraphics[width=\linewidth]{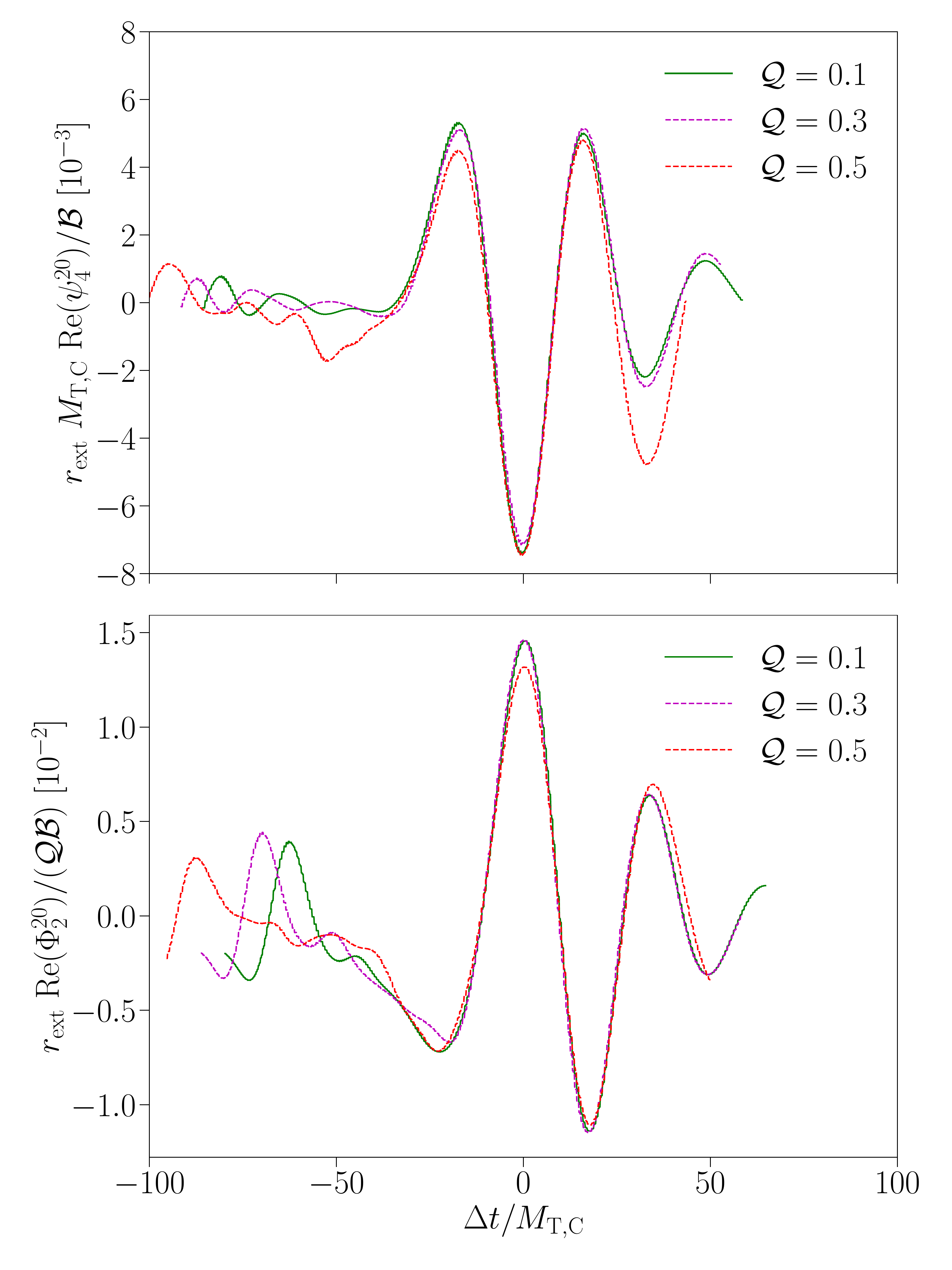}
\caption{\label{fig:HADwaveforms1} The $l=2, m=0$ mode (in the rotated frame with the black holes on the $z$-axis) of $\psi_4$ (top panel) and $\Phi_2$ 
(bottom panel) aligned at merger using the peak of the amplitude of the real part of the waveform. These waveforms correspond to head-on collisions of equal-mass, nonspinning charged black holes evolved using HAD for different charge to mass ratios, i.e., \texttt{ho-v0qp1}, \texttt{ho-v0qp3}, and \texttt{ho-v0qp5}.}
\end{figure}

In Fig.~\ref{fig:HADwaveforms1}, we show the $l=2, m =0$ mode of gravitational and electromagnetic radiation from the
head-on collisions of binary black holes for various charge-to-mass ratios, including the $\mathcal{B}:= 1 -
\mathcal{Q}^2$ Newtonian scaling that~\cite{Zilhao:2012gp} found to account for most of the amplitude's dependence on the charge.
In~\cite{Zilhao:2012gp} the authors place the black holes on the $z$-axis, whereas in our setup, we place them on the
$x$-axis. Thus, to make a direct comparison with~\cite{Zilhao:2012gp}, we transformed our waveforms to place the black
holes along the $z$-axis using the {\sc{quaternionic}} package~\cite{Boyle:2021quaternionic,Boyle:2016tjj}. We find that
the waveforms from our evolutions satisfy the scalings found in~\cite{Zilhao:2012gp} for the merger-ringdown portion of
the waveform with reasonable accuracy, though with larger differences than in~\cite{Zilhao:2012gp}. Specifically, \cite{Zilhao:2012gp} found differences of at most $2\%$ compared to the
scaling for charges up to $98\%$ of maximal. In our simulations, however, we found the differences to be larger (up to~$\sim 3\%$)
for charges up to $50\%$ of maximal, though our evolutions are preliminary, and we did not attempt to assess convergence, in addition to the issues
with the effects of the filling mentioned previously.

\subsubsection{Head-on collisions of boosted uncharged binary black holes}

We also considered head-on collisions of uncharged black holes boosted towards each other with velocity parameters of $0.1$ and $0.3$, as well as an unboosted case, for comparison. In the boosted cases, the assumption that the system is in quasi-equilibrium breaks down, and hence our choice of setting $u_{ij} = 0$ becomes increasingly inaccurate for larger initial velocities. However, we still find that increasing the initial velocity decreases the time to merger, so we present these results as a further example of the code's ability to generate initial data for generic binaries. 

To quantify the effects of the boost, we consider the time to merger for each of the simulations ($t_{\rm{sim}}$), which we compare with the Newtonian estimate ($t_{\rm{m}}$), as in the charged case considered above. To compute the merger time from the simulations, we use the time it takes for each black hole to reach the origin, using the minimum of the lapse as a proxy for the location of the black hole. To compute the Newtonian estimate for the merger time, we use energy conservation. Specifically, we generalized the calculation done in~\cite{Zilhao:2012gp} to include boosts. For two black holes each with mass $M/2$ and charge $Q/2$ placed initially at $x = \pm d/2$ with initial velocities of $v_0$, conservation of energy implies 
\begin{equation}
	M \dot{x}^2 - \dfrac{M^2 \mathcal{B}}{4x} = M v_0^2 -\dfrac{M^2 \mathcal{B}}{2d},
\end{equation}
where $x$ is the position of one of the black holes and the overdot denotes a time derivative.
Rearranging and integrating both sides, we get
\begin{align}
	t_{\text{m}} = \dfrac{\mathcal{B} M}{\mathcal{Z}^{3}}\left[\dfrac{\pi}{2} -  \arctan\left(\dfrac{2v_0} {\mathcal{Z}}\right)\right] -\dfrac{2v_0 d}{\mathcal{Z}^{2}},
\end{align}
where
\begin{equation}
	\mathcal{Z} := \sqrt{\dfrac{\mathcal{B}M}{d} - 4 v_0^2},
\end{equation}
where we evaluate the arctangent on the right-hand side of the branch cut (so its real part is $\pi/2$) for the cases where $\mathcal{Z}$ is imaginary, which ensures the result is always real. 
For the unboosted case (\texttt{ho-v0q0}), we find the time from the simulation, $t_{\rm{sim}}$, to be larger than $t_{\rm{m}}$ with a relative error
$\sim 18\%$. For the boosted cases, similar to the charged case, we scale both $t_{\rm{sim}}$ and $t_{\rm{m}}$ by the
corresponding computation for the unboosted case and denote these scaled values by $\tilde{t}_{\rm{m}}$ and $\tilde{t}_{\rm{sim}}$,
respectively. We find $\tilde{t}_{\rm{sim}}$ to be larger than $\tilde{t}_{\rm{m}}$ for both \texttt{ho-vp1q0} and
\texttt{h0-vp3q0}, and the relative error between them to be $\sim 57\%$ for \texttt{ho-vp1q0}, and $\sim
103\%$ for \texttt{h0-vp3q0}.

There could be several reasons behind these large discrepancies for both charged and uncharged, boosted cases. In particular, our method to infer the collision time from the numerical evolutions likely overestimates the merger time compared, e.g., to using the first appearance of a common apparent horizon to determine the merger time, as was done in~\cite{Zilhao:2012gp}. Additionally, for the boosted cases, as mentioned before, we use the quasi-equilibrium approximation ($u_{ij} = 0$) beyond its regime of validity. This, in addition to the issues with filling the black holes described earlier and the effects of junk radiation, can result in the black holes in the simulation having different initial velocities than those used in the construction of the initial data. 

As a rough check, we estimated the initial velocities of the black holes using the location of the black hole and finite differencing. We started at $13.5 M_{\rm{T},\rm{C}}$ to be after the initial transients and considered a time span of $7 M_{\rm{T},\rm{C}}$, chosen by eye so that a constant velocity given by the finite difference gives a good approximation to the time dependence of the black holes' positions. We found the initial velocities to be $0.087$ and $0.130$ for \texttt{ho-vp1q0} and \texttt{h0-vp3q0}, respectively. For \texttt{h0-v0q0}, we found the computed velocity to be zero, in agreement with the velocity used in the computation of the initial data. Using these velocities in the merger time calculation reduced the computed merger time as well as the relative error between $\tilde{t}_{\rm{m}}$  and $\tilde{t}_{\rm{sim}}$ to $\sim 23\%$ and $\sim 74\%$ for \texttt{h0-vp1q0} and \texttt{h0-vp3q0}, respectively.

\section{Summary and Future Work}
\label{sec:discussions} 
In this paper, we constructed initial data for spinning, charged, orbiting binary black
holes by extending the superposed Kerr-Schild construction for vacuum binary black hole initial data
presented in~\cite{Lovelace:2008hd}. In
contrast to previous work~\cite{Bozzola:2019aaw} which constructed conformally flat
puncture initial data, our construction gives conformally curved initial data with
excision. Our approach---in addition to providing a completely independent method to
construct charged binary black hole initial data---offers several advantages over the
approach used in~\cite{Bozzola:2019aaw}, e.g., better
control over the physics through the boundary conditions at the excision surface and the
ability to construct binary initial data with larger spin, above the Bowen-York limit.
It also provides a complementary implementation of the original vacuum construction
in~\cite{Lovelace:2008hd} with a different numerical setup and a slightly different
choice of boundary conditions. Specifically, we transfer the corotation and expansion terms from
the outer to the inner boundary condition for the shift. 

We tested our initial data construction for several cases including binaries with highly charged black holes and black holes with both charge and spin. We performed convergence tests and found our
initial data implementation gives exponential convergence for low resolutions, but slower convergence for higher resolutions. We also carried
out exploratory evolutions of charged and uncharged initial data using two different evolution codes (BAM in the uncharged case and HAD with both charged and uncharged data). Here we evolved a
quasicircular uncharged binary without charge with BAM, but were unable to obtain an orbiting evolution with HAD even in the uncharged case, which we attribute to the differences in filling inside the excision surfaces in the two codes.
Thus, in the charged case, we considered head-on collisions with different charge-to-mass
ratios and found that our numerical results are consistent with the simple (Newtonian) analytic scalings presented in~\cite{Zilhao:2012gp}. Our estimates of the time of collapse for boosted, head-on collisions differed significantly from a similar Newtonian estimate. However, a number of factors particular to these results might explain such differences, and the collapse time decreased with increasing initial velocity as would be expected.

There are several improvements one could make to our initial data construction. For example, it would be useful to
understand and mitigate the loss of exponential convergence at higher resolutions, especially for binaries with large
charge. This might require conformally rescaling the EM quantities as was done in~\cite{Alcubierre:2009ij,
Bozzola:2019aaw}, and careful handling of the regularity conditions at the computational domain boundaries, where the
gradients of electric constraint violations are the largest. In order to generate waveforms for gravitational wave data
analysis, we might need a better initial guess for the eccentricity reduction procedure that does not lead to the unusual eccentricity reduced solution we obtained here. We will also need to resolve the outstanding issues with black hole filling in HAD. In particular, it would be worth exploring if setting the excision surface to be inside the apparent horizon, as in~\cite{Varma:2018sqd}, mitigates the issues we encountered while evolving our excision initial data using puncture methods. 

As discussed in Sec.~\ref{sec:intro}, one goal is to use our initial data to compute low-eccentricity charged binary black
hole waveforms that are sufficiently accurate for data analysis applications. Such waveforms will allow us to check how
sensitive current LIGO-Virgo tests~\cite{LIGOScientific:2020tif,LIGOScientific:2021sio} are to completely consistent, parameterized deviations
from GR. To evolve our charged initial data, another possibility---in addition to HAD---would be the LEAN
code~\cite{Sperhake:2006cy, Zilhao:2013nda, Zilhao:2012gp} which would allow us to compute the quasilocal mass, charge,
and spin of black holes during evolution.

Our construction can also be extended to construct initial data for binary black
holes in Einstein-Maxwell-dilaton theory. In general, this would require superposing numerically constructed single black
hole solutions for spinning black holes~\cite{Kleihaus:2002tc,Kleihaus:2003df}, but analytical solutions are known for
black holes without spin and for spinning black holes with specific values of the coupling parameter~\cite{Frolov:1987rj},
so an initial implementation could focus on these simpler cases. There are no additional constraint equations to solve for
the dilaton field, just additional source terms in the Hamiltonian and momentum constraints. However, one might ultimately want to generalize the construction to also allow the dilaton field to be adjusted as part of the solve, to obtain a better
approximation to quasiequilibrium. Evolutions with such initial data would improve upon the existing work by
Hirschmann~\emph{et al.}~\cite{Hirschmann:2017psw} using approximate initial data, and also allow for comparisons with
analytical work~\cite{Khalil:2018aaj,Julie:2018lfp}.

\acknowledgments
We thank Erik Schnetter for useful discussions and comments on the manuscript.
Additionally, we thank Geoffrey Lovelace and Ulrich Sperhake for useful discussions. This
research was supported in part by Perimeter Institute for Theoretical Physics. Research
at Perimeter Institute is supported in part by the Government of Canada through the
Department of Innovation, Science and Economic Development Canada and by the Province of
Ontario through the Ministry of Colleges and Universities. S.~M.\ also acknowledges the
hospitality of the International Center for Theoretical Sciences, Tata Institute of
Fundamental Research (ICTS-TIFR), where part of this work was conducted. N.~K.~J.-M.\
acknowledges support from the AIRBUS Group Corporate Foundation through a chair in
``Mathematics of Complex Systems'' at ICTS-TIFR and STFC Consolidator Grant
No.~ST/L000636/1. W.~T.\ was supported by the National Science Foundation under grants
PHY-1707227 and PHY-2011729. S.~L.~L.\ was supported by the NSF under grants PHY-1912769
and PHY-2011383. Computations were performed on Symmetry at Perimeter, Cosmos at DAMTP,
Cambridge (funded by BEIS National E-infrastructure capital grants ST/J005673/1 and STFC
grants ST/H008586/1, ST/K00333X/1), the Cambridge Service for Data Driven Discovery
(CSD3) system (through DiRAC RAC-13 Grant No.~ACTP186), Maple at the University of
Mississippi (funded by NSF Grant CHE-1338056), and XSEDE resources.

\appendix
\section{Kerr-Newman black hole in Kerr-Schild form}
\label{append:KS}
In Sec.~\ref{sec:id}, we discuss the superposition of Kerr-Newman metrics for the choice of freely specifiable variables. Here, we give the relevant expressions for completeness. In the $3+1$ setting, the spatial metric and extrinsic curvature for a stationary Kerr-Newman black hole in Kerr-Schild coordinates are given by~\cite{Debney:1969zz, Newman:2014}
\begin{subequations}
	\label{eq:gammaK}
\begin{eqnarray}
	\gamma_{ij} &=& f_{ij} + 2H l_i l_j, \\
	K_{ij} &=& \dfrac{1}{2\alpha} \left(\nabla_i\beta_j + \nabla_j\beta_i - \partial_t \gamma_{ij}\right),
\end{eqnarray}
\end{subequations}
where the lapse $\alpha$ and the shift $\beta_i$ are given by
\begin{equation}
	\label{eq:alphabeta}
	\alpha = \dfrac{1}{\sqrt{1+2Hl_{0}^{2}}}, \qquad \beta_{i} = 2Hl_{0}l_{i},
\end{equation}
writing these in a form that makes it easy to compute their boosted form, following~\cite{Bonning:2003im}.
In the above equations, the metric function $H$ and the null vector $l_\mu \doteq \{l_{0}, l_i\}$ are given by
\begin{subequations}
\begin{align}	
	H  = {} & \dfrac{Mr^{3} - (Qr)^{2}/2}{r^{4} + \left(\mathbf{a}\cdot\mathbf{x}\right)^{2}}, \\
	\label{eq:la}
	l_\mu \doteq {} & \Big\{1, \dfrac{r\left[\mathbf{x} - \left(\mathbf{\hat{a}}\cdot\mathbf{x}\right)\mathbf{\hat{a}} \right] 
	- \left(\mathbf{a}\times\mathbf{x}\right)}{r^2 + a^2} + 	\dfrac{\left(\mathbf{\hat{a}}\cdot\mathbf{x}\right)\mathbf{\hat{a}}}{r}\Big\},
\end{align}
\end{subequations}
where $\mathbf{x}\doteq\{x,y,z\}$ is the three dimensional coordinate vector, $\mathbf{a} = a\,\hat{\mathbf{a}}$ is the angular momentum per unit mass, and $M$ and $Q$ are the mass and charge of the black hole, respectively. Here we use $\doteq$ to denote that we are giving the components of a vector. 
The parameter $r$ is given by
\begin{subequations}
\begin{align}
	r^{2} &= \dfrac{1}{2}(\rho^2 - a^2) + \sqrt{ \dfrac{1}{4}\left(\rho^2 - a^2\right)^2
	+ \left(\mathbf{a}\cdot\mathbf{x}\right)^{2}}, \\
	\rho^2 &= x^2 + y^2 + z^2.
\end{align}
\end{subequations}
Note that in Eq.~\eqref{eq:la} and in Eq.~\eqref{eq:vector_potential_rotated} below, we use the Euclidean cross product. To compute the electric and magnetic fields for a Kerr-Newman black hole, we use the four-potential~\cite{Newman:2014}
\begin{align}
\label{eq:vector_potential_rotated}
A_\mu \doteq -\dfrac{Qr^{3}}{r^{4} + \left(\mathbf{a}\cdot\mathbf{x}\right)^{2}} {} &  \Big\{1, \dfrac{r\left[\mathbf{x} -
\left(\hat{\mathbf{a}}\cdot\mathbf{x}\right)\hat{\mathbf{a}} \right]
- \mathbf{a}\times\mathbf{x}}{r^2 + a^2} \nonumber\\ 
{} & + \dfrac{\left(\hat{\mathbf{a}}\cdot\mathbf{x}\right)\hat{\mathbf{a}}}{r}\Big\},
\end{align}
and compute the electric and magnetic field using~\cite{Alcubierre:2009ij}
\begin{equation}
	\label{eq:EM}
	E^{\mu} := -n_{\nu}F^{\nu\mu}, \quad B^{\mu} := -n_{\nu}{}^*F^{\nu\mu},
\end{equation}
where $n_\mu \doteq \left\{\alpha, 0,0,0\right\}$ is the four velocity of an Eulerian observer, and $F_{\mu\nu}$ and ${}^*F^{\mu\nu}$ are the Faraday tensor and its dual
\begin{equation}
	\label{eq:Faraday}
	F_{\mu\nu} = \partial_{\mu}A_{\nu} -  \partial_{\nu}A_{\mu}, \quad {}^*F^{\mu\nu} := -\dfrac{1}{2}\epsilon^{\mu\nu\eta\gamma}F_{\eta\gamma}.
\end{equation}
Here $\epsilon^{\mu\nu\eta\gamma}$ is the curved-space Levi-Civita symbol, such that $\epsilon^{0123} = -1/\sqrt{-g}$, where $g$ is the determinant of the full 4-dimensional spacetime metric. 

For a boosted Kerr-Newman black hole, we first compute $H$ and the vector quantities $U_\mu\in\{l_\mu, n_\mu, A_{\mu}\}$ in the boosted coordinates $\bar{x}_\alpha$, given by 
\begin{equation}
		x^{\beta} = \Lambda^{\beta}_{~~\alpha}\,\bar{x}^{\alpha},
\end{equation}
where $x^{\alpha}$ are the inertial (grid) coordinates and $\Lambda^{\beta}_{~~\alpha}$ is the Lorentz transformation matrix relating the two frames, and then apply the Lorentz transformation to the vectors themselves, i.e., 
\begin{subequations}
\begin{align}
	H(x^{\alpha}) = {}  &  \bar{H}\bigl([\Lambda^{-1}]^{\beta}_{~~\alpha}\,x^{\alpha}\bigr), \\
	U_{\delta}(x^{\alpha})  = {} &  \Lambda^{\gamma}_{~~\delta}\, \bar{U}_{\gamma}\bigl([\Lambda^{-1}]^{\beta}_{~~\alpha}\,x^{\alpha}\bigr),
\end{align}
\end{subequations}
where barred quantities correspond to the boosted frame.
We then use Eqs.~\eqref{eq:gammaK},~\eqref{eq:alphabeta},~\eqref{eq:EM}, and~\eqref{eq:Faraday} to compute the relevant quantities for the boosted black hole.

\section{Filling inside excision surfaces in BAM}
\label{append:BHFiller}
To fill inside the excision surfaces in BAM, we modified the
original BHFiller algorithm (see Chapter 3.2 in~\cite{Reifenberger:2013th} for details) that fills inside spherical excision surfaces to allow for our excision surfaces that are deformed by the black holes' spin and boosts. Specifically, we first transform from the grid coordinates to coordinates in which the excision surfaces are spherical, then perform the interpolation using BHFiller, and finally transform back to grid coordinates. 

The coordinate transformation to coordinates in which the excision surfaces are spherical is a simple linear one and is performed in two steps, first removing the Lorentz contraction due to the boost and then removing the deformation due to the spin. These transformations are both of the form
\begin{equation}
	\mathbf{x} \to w_\perp\mathbf{x} + (w_\parallel - w_\perp)(\mathbf{x}\cdot\hat{\mathbf{k}})\,\hat{\mathbf{k}},
\end{equation}
where the parameters $w_\parallel$ and $w_\perp$ set the amount of deformation parallel and perpendicular to the symmetry axis given by the unit vector $\hat{\mathbf{k}}$. Specifically, for the Lorentz boost, with velocity $\mathbf{v}$ (and magnitude $v$), one takes $w_\parallel = 1/\sqrt{1 - v^2}$, $w_\perp = 1$, and $\hat{\mathbf{k}} = \hat{\mathbf{v}}$. For the spin deformation, with angular momentum per unit mass of the black hole $\mathbf{a}$ (and magnitude $a$), one takes $w_\parallel = 1$, $w_\perp = r_+/\sqrt{r_+^2 + a^2}$, and $\hat{\mathbf{k}} = \hat{\mathbf{a}}$, where $r_+$ is the coordinate radius of the Kerr outer horizon, given (for Kerr-Newman) in Eq.~\eqref{eq:r_plus}. (Here we only consider Kerr because BAM only evolves the vacuum case.) The inverse transformations are given by using the reciprocals of $w_\parallel$ and $w_\perp$.

\section{Filling inside excision surfaces in HAD}
\label{append:BHFillerHAD}
To fill the black hole interiors in HAD, we use an approach seeking to approximate the BHFiller method described in~\cite{Reifenberger:2013th}. Using the mask provided by SGRID, we fill only those
points flagged as ``excised'' to avoid throwing away any information from the initial data solver. In particular, the filling algorithm uses four passes through the domain. The first three
passes each correspond to one of the coordinate directions ($x$, $y$, and $z$) during which each interior point is updated with a sweep from each side of the hole.
For a black hole centered at $(x_c,y_c,z_c)$, the sweep starts at small $x$ and fixed $y$ and $z$. When an excised point is first encountered, the value of the field $u$ at the last un-excised point, $u(x_0,y,z)$
together with a backward difference estimate of the derivative $\Delta u/\Delta x$ are used  
to extrapolate points as the sweep proceeds towards $x_c$. Similar sweeps in the $y$ and $z$ directions follow as
\begin{subequations}
\begin{align}
   {}^1 u(x,y,z) = {} & u(x_0,y,z)+ \left(x-x_0\right) \frac{\Delta u}{\Delta x} \left( \frac{x-x_c}{r_A} \right)^2, \\
   {}^2 u(x,y,z) = {} & {}^1 u(x,y_0,z)  + \left(y-y_0\right) \frac{\Delta u}{\Delta y} \left(\frac{y-y_c}{r_A} \right)^2, \\
   {}^3 u(x,y,z) = {} & {}^2 u(x,y,z_0)  + \left(z-z_0\right) \frac{\Delta u}{\Delta z} \left( \frac{z-z_c}{r_A} \right)^2.
\end{align}
\end{subequations}
This expression is for excision surface $A$; recall that $r_A$ is the coordinate distance from the center of that excision surface.
These passes serve to approximate the linear extrapolation from the exterior to the center of the black hole along the radial direction. The final pass 
\begin{equation}
   {}^4 u(x,y,z) = \Gamma \; {}^3 u(x,y,z)  + \left( 1-\Gamma\right) u(x_c,y_c,z_c),
\end{equation}
interpolates the extrapolated value with a prescribed central value for the field. This last step is described in~\cite{Reifenberger:2013th} which gives both the form of the interpolating function $\Gamma$ as a function of distance from the center of the black hole and the central values for each
of the fields filled. For the electric and magnetic field components, we set the central value to zero.

\bibliography{ChargedBBH}
\end{document}